\documentclass[fleqn,10pt]{wlscirep}
\usepackage[utf8]{inputenc}
\usepackage[T1]{fontenc}
\usepackage{comment}
\usepackage{subfig}

\title{Prediction and visualization of Mergers and Acquisitions using Economic Complexity}

\author[1,5]{Lorenzo Arsini}
\author[1,2,4,*]{Matteo Straccamore}
\author[3,2]{Andrea Zaccaria}

\affil[1]{Dipartimento di Fisica, Universit\`a ``Sapienza”, P.le A. Moro, 2, 00185 Rome, Italy}
\affil[2]{Centro Ricerche Enrico Fermi, Piazza del Viminale, 1, 00184 Rome, Italy}
\affil[3]{Istituto dei Sistemi Complessi (ISC) - CNR, UoS Sapienza,P.le A. Moro, 2, 00185 Rome, Italy}
\affil[4]{SONY Computer Science Laboratories, Paris, 6, rue Amyot, 75005, Paris, France}
\affil[5]{INFN, Section of Rome, Rome, Italy}

\affil[*]{matteo.straccamore@cref.it}

\begin{abstract}
Mergers and Acquisitions represent important forms of business deals, both because of the volumes involved in the transactions and because of the role of the innovation activity of companies. Nevertheless, Economic Complexity methods have not been applied to the study of this field.
By considering the patent activity of about one thousand companies, we develop a method to predict future acquisitions by assuming that companies deal more frequently with technologically related ones. We address both the problem of predicting a pair of companies for a future deal and that of finding a target company given an acquirer. We compare different forecasting methodologies, including machine learning and network-based algorithms, showing that a simple angular distance with the addition of the industry sector information outperforms the other approaches. Finally, we present the Continuous Company Space, a two-dimensional representation of firms to visualize their technological proximity and possible deals. Companies and policymakers can use this approach to identify companies most likely to pursue deals or to explore possible innovation strategies.
\end{abstract}

\begin{document}

\flushbottom
\maketitle

    \thispagestyle{empty}

\section*{Introduction}
Mergers \& Acquisitions (M\&A) are one of the most popular forms of business development, and represent the subject of considerable research in financial economics  \cite{bruner2004applied}.
Such operations are used extensively as a financial instrument by firms of any region and size and constitute a business that, only in 2019, has almost reached 4 trillion dollars (Source: Institute for Mergers, Acquisitions and Alliances (IMAA) \url{https://imaa-institute.org}).
Despite the huge spread of the phenomenon, from a statistical point of view it is recorded that, on average, a M\&A does not bring significant economic benefit to the involved companies \cite{King2004meta}. There is, however, a strong variance in the data, that includes both acquisitions of huge success as well as dramatic failures. Despite the various attempts, there is no agreement in existing academic research about the right variables that decisively influence the realization and the outcome of an acquisition \cite{Gomes2013critical, Ismail2009Review}. The complexity of the phenomenon under consideration, as an economic and social process, is reflected in the heterogeneity of the studies carried out in this field, which lack comprehensive theoretical models and common variables \cite{Rossi2013review}. 
In this work, we study M\&A using tools and data from the Economic Complexity framework and, in particular, the concept of Relatedness \cite{hidalgo2018principle,Hidalgo2007,zaccaria2014taxonomy}, that we use to compare the patenting activity of companies. \\
Every year, hundreds of new technologies are developed, as processes and products become increasingly complex.
In such a rapidly changing environment, it is crucial for a company to stay ahead and get a good position in the innovation race.
Often, developing new technologies in internal R\&D environments is not enough or convenient in terms of time and costs.
Because of this, many companies seek to expand their technological horizon by undergoing an acquisition.
Through a technological acquisition, the acquiring company can absorb the target’s capabilities, recombine technologies to produce innovation, intensify internal research and skills development \cite{Cefis2010impact} or launch products into a new market.
In these terms, an acquisition can be seen as an expansion of the acquirer's knowledge and capability base. This kind of subject has been widely addressed in the corporate strategy literature that focuses on firms' diversification, as discussed in the following. Here we restrict our study to deals in which it is possible to individuate an acquirer and a target company. We refer to this set of deals with the generic term "acquisitions".\\
Several efforts have been done in researching a comprehensive theory on the diversification of firms (both at productive and technological level) since the early studies of Penrose and Teece \cite{penrose1959, TEECE1982}.
Many authors took up these ideas to explain the reasons under the diversification of firms, the way in which they expand and the financial outcome (for a complete literature review we refer to \cite{knecht2013}).
Beyond the concept of diversification as the simple scope of (both productive and patenting) activities in which a firm is involved, recent research focused on the key concept of the \textit{relatedness} between these activities.
A seminal work on this subject is the one of Teece et al. \cite{Teece1994understandig}. The authors measure the similarity between activities by counting the excess of their co-occurrences with respect to a suitable null model. Teece et al. show that activities that are more related are also more frequently combined within the same company. %Starting from this hypothesis, they build a relatedness measure between companies and the 4-digit SIC industrial codes relative to the industrial sectors in which they operate, %showing that firms tend to diversify in activities that are more related to the ones they are already involved in.
%The survivor principle and the
Teece's relatedness metrics were taken up by several authors whose researches focus on the internal coherence of firms' technological portfolios, that is, the innovative sectors to which their patents belong.
This stream of literature substantially confirm Teece's results also in the technological field \cite{Breschi2003knowledge, Piscitiello2000relatedness, Bottazzi2010measuring}.
The patent-based approach%, employed also in this work,
offers various advantages such as the wide geographical coverage and the richness of information that patents provides about inventions (e.g. bibliographic data, citations, claims, technological fields impacted by the patents, etc.) \cite{ERNST2003233, Strumsky}.
Similar themes were also faced using the approach known as Economic Complexity.
This approach stands out for the use of tools from complexity science, such as co-occurrence networks \cite{Hidalgo2007, HidalgoK2007} and machine learning algorithms \cite{tacchella2021relatedness,albora2021product}.
In \cite{pugliese2019coherent} this approach is used to study the technological diversification of firms and the relatedness between their activities. The authors introduced the concept of \textit{coherent diversification} and showed that firms that diversify their patenting activity in a coherent manner, i.e. expanding in related sectors, have on average higher levels of labor productivity. 
Relatedness measures between companies and technological sectors can also be used to forecast the technological and productive diversification of firms \cite{straccamore2022will,albora2022machine}.
In this paper we argue that methods similar to the ones described in these papers can be used to build relatedness measures to study the phenomenon of Mergers and Acquisitions (M\&A).\\
Despite some sort of heterogeneity in the M\&A economic literature, it is still possible to identify some principles and recurring ideas.
First of all, the concept of \textit{absorptive capacity} \cite{Cohen1990AbsorbCap}, as the ability of the acquirer company to assimilate knowledge and competencies of the target firm. Extending this concept, good integration between acquirer and target companies is thought to be linked to the relatedness between the two \cite{Lane1998relative}.
The concept of \textit{relatedness}, and in general the interaction between resources and capabilities of the companies involved in a M\&A process, is definitely the most widespread in the literature and is applied in many different contexts.
For example, it has been shown how geographical distance negatively influences the probability of a M\&A to occur \cite{Kaul2015ACP, Chakrabarti2016therole}, and also how similarities in terms of the ownership or the industrial sector can, on the contrary, increase such probability \cite{Bettinazzi2020OwnershipSI, Kennedy2002matching}.
In \cite{cefis2013theimportance}, analyzing a large set of acquisitions and employing a similarity measure introduced by Teece et al. in \cite{Teece1994understandig}, a statistically significant correlation between the occurrence of a M\&A and the \textit{industry relatedness} is found.
As in the economic literature that focuses on diversification, in the last two decades a section of the M\&A studies started to analyze acquisitions from the point of view of the patenting activity of involved firms.
This stream of literature focused on the similarity between the companies' \textit{knowledge bases}, or in other words, their \textit{technological relatedness}. 
One of the first studies of this kind is the one of Ahuja-Katilia \cite{AhujaKatilia2001}. The authors computed a measure of technological similarity between the acquirer and the target firms as the overlap of the set of all patents produced and cited by those firms. This measure is found to have an inverse parabolic behavior with the innovation performance after the acquisition. In other words, the optimal post-acquisition performances are observed for an intermediate level of technological similarity.
In a similar fashion, many successive authors built different measures of technological relatedness and applied them to companies involved in M\&A processes and tried to link them to post-acquisition performances.
Examples of this stream of literature can be found in the work of Cloodt et al. \cite{Cloodt2006}, Cassiman et al. \cite{CASSIMAN2005195} and many others \cite{Hagedoorn2002, valentini2010aa, Jo:2016aa, makri2010, orsi2015}.
Although several studies recur the idea of this inverse parabolic behavior between relatedness and performances, results are not yet conclusive. Indeed for now, there is not yet a standard, recognized and effective method to build robust performances \cite{Jo:2016aa} or relatedness measures \cite{cimini2022meta}, and as shown in \cite{Jo:2016aa}, results may vary on metrics definitions. 
In general, the majority of the M\&A literature that builds relatedness measures between acquirers and target firms focuses on correlating such measures with successive performances and not on using them for predictions. However, as pointed out in \cite{albora2021product}, we believe that a forecast constitute an important test to compare the goodness of relatedness assessments. Notable forecast exercise includes \cite{wei2008patent}, in which an ensemble learning algorithm is trained on a set of relative features between companies, built using patent data, to predict future acquisitions, and the attempt to M\&A prediction in \cite{FUTAGAMI202122}.
In this latter work, a very large set of M\&A features is built employing financial, geographical, industrial and patent data of firms. Then a tree-based algorithm is trained for M\&A prediction.

In short, although the importance of relatedness between acquirer and target in an acquisition is now recognised, there is not yet a standard method for calculating and, most importantly, evaluating the goodness of such relatedness measures. There is, also, a fundamental lack of studies that compare different relatedness measures in a systematical way.
In this heterogeneous context we propose our M\&A prediction study. We build upon a capability view of firms and follow the methods and ideas of the Economic Complexity stream of literature. Starting from patent data we define different relatedness measures between firms to exploit their technological affinity.
These metrics are the used to make predictions on possible M\&A pairs of companies and results are evaluated as in a machine learning classification problem.
In this way we can study in a systematical way the degree in which the considered relatedness measure are able to discriminate between pair of companies that complete an acquisition and randomly assembled pairs. We find that both the technological and the industrial affinity play a role, and that cosine similarities outperform the others.
Finally, we believe that it is also fundamental to have a simple visual representation of results for a straightforward interpretation, an effort lacking in the present literature.
To fill this gap, we adapt the concept of Continuous Projection Space \cite{tacchella2021relatedness,straccamore2022will} to our case, building a 2D space in which related companies are close to each other.

%\textcolor{red}{in conclusione, quali sono i problemi aperti in letteratura? problemi noti e problemi dei quali ci siamo accorti noi}

%\textcolor{red}{hypothesis: similar firms on average make M\&A. è corretto? trova riscontro in letteratura?}\\
%\textcolor{red}{dovremmo dire meglio in cosa ci differenziamo rispetto alla letteratura e perché queste differenze sono importanti. forecast come test per fare confronti in maniera sistematica- visualizzazione - confronto relatedness sect-comp vs sect-sect vs comp-comp - prevediamo cose diverse - aggiungete altro}

% Results and Discussion can be combined.
\section*{Results}
Our investigation consists of four steps: i) we compute different measures that quantify how much a company, possible target of an acquisition, is related to the present technological activity of the acquirer company; ii) we use these relatedness measures to forecast whether a deal will happen or not; iii) we quantify our ability to forecast the deals using different relatedness and performance measures; iv) we represent the deals in a two-dimensional plane. Before discussing our findings, we briefly introduce our methodology and data. More details are provided in the Methods section.
\\
\subsection*{Data and testing procedure}
The results of this work are based on the patent data coming from the PATSTAT database (\url{www.epo.org/searching-for-patents/business/patstat}) that contains information about over $40000$ patents and their technology sectors of belonging. These are classified by the use of technology codes that are encoded using IPC's 6 digits classification (\url{https://www.wipo.int/classifications/ipc/en/}). From now on we will refer to these technology codes as ``technologies". Subsequently, this information is matched with the AMADEUS database (\url{https://amadeus.bvdinfo.com}), which covers over 20 million firms with European registered offices. 
Finally, the M\&A information comes from two different databases: Crunchbase (\url{https://www.crunchbase.com}) and Zephyr (\url{https://login.bvdinfo.com/R0/zephyrneo}).
The final set of deals used for the analyses presented in this paper is made up of $8737$ companies of which $913$ are involved in $547$ M\&A deals. We select these companies because we can assign to them a univocal industrial sector, based on Crunchbase data on industrial sectors.
Complete information on data processing and Crunchbase name matching, and industrial sectors classification can be found in the Methods section and in the Supplementary Information.
With these data, we can associate patents, and more specifically the technologies, to the companies involved in M\&A; in particular, we build a temporal bipartite networks connecting patenting companies with their technologies.\\
This temporal network is represented by $13$ yearly adjacency matrices $\textbf{M}^y$, one for each year $y$ from 2000 to 2012 that link $8737$ companies to $7132$ technologies. The matrix element $M_{ft}^y$ represents how much a technology $t$ is present in the patenting activity of firm f in year $y$: specifically, given a year $y$, we assign to each patent one unit of weight; this is then divided into equal shares between all the observed (firm f-technology t) pairs and, finally, the matrix is built by summing element-wise these contributions. This procedure takes into account that, usually, more than one code is present in each patent and rarely a single patent is submitted by more than one applicant firm.\\
Moreover, assuming that a patent filed in a certain year is representative of the firm’s capabilities also in the following years, we will also consider the matrices $M_{ft}^Y$, each defined as the sum of $M_{ft}^y$ over the years from 2000 to $Y$.\\
The $\textbf{M}^Y$ matrices can be used to train different algorithms to calculate our predictions about possible M\&A. In particular, we use $\textbf{M}^Y$ to calculate the similarity between each pair of companies; such similarity is assumed to be related to the probability that two companies will have a M\&A in the year $Y$.

 \subsection*{Measures of similarity between companies}
Our predictions of M\&A events are based on various measures of business affinity, based using only patent data with adding in some cases information related to the company's industry sector. We give here a brief description of these metrics, referring the interested reader to the Methods section for a more detailed explanation. We will call a metric of \textit{similarity} if it is computed between elements of the same type, for example between two companies, or two technologies. Instead, we will refer to metrics of \textit{relatedness} if they are computed between different elements, such as a company and a technology. We divide our metrics into three different categories:
\begin{itemize}
    \item Direct measures. These metrics are based on the construction of a similarity measure between firms. In this case, we can think of firms as vectors in a technology codes' space with coordinates $(M^Y_{f1}, ..., M^Y_{ft}, ... , M^Y_{fn})$, where $n = 7132$ is the number of technologies, that is the dimensionality of the space in which the firm-vectors are defined.
    We use as \textit{direct} measures: 
    \begin{itemize}
        \item \textbf{Common Tech}, the scalar product between two firm-vectors, which provides the number of technologies that co-occur in both companies, possibly fractional, since $\textbf{M}$ elements can be fractional;
        \item \textbf{Jaffe}, the cosine of the angle between two firm-vectors, introduced by Adam Jaffe in \cite{Jaffe1986technological} and adopted in this context in \cite{valentini2010aa};
        \item \textbf{Euclidean Distance (EU)}: the inverse euclidean distance between two firms;
        \item \textbf{Jaffe + Sectors (J+S)}: we incorporate the information on technological portfolios (given by the Jaffe measure) with the information regarding the companies' industrial sector. In formula:
        \begin{equation*}
            P_{ff'} = \alpha S_{ff'} + (1-\alpha) J_{ff'},
        \end{equation*}
        where $J_{ff'}$ represents the Jaffe measure between the firms' technological portfolio, $S_{ff'}$ is equal to $1$ if both firms $f$ and $f'$ belong to the same sector and equal to $0$ otherwise, and finally the parameter $\alpha$ can be tuned according to the goodness of the prediction on M\&A processes. We find that the optimal value of $\alpha = \hat\alpha$ depends on the class imbalance of the prediction exercise; a detailed investigation is presented in the Methods section.
    \end{itemize}
%and \textbf{Jaffe + Sectors}, which needs a more detailed description. This measure incorporates the information on technological portfolios (given by the Jaffe measure) with the information regarding the companies' industrial sector, given by our classification described in the Methods section. This measure of similarity between firms consists in a linear combination of two different quantities:
%\begin{itemize}
    %\item{\textbf{$J_{ff'}$}: the cosine similarity (or Jaffe measure) between the firms' technological portfolio,}
    %\item{\textbf{$S_{ff'}$}: a value that is equal to 1 if both firms belong to the same sector and equal to 0 otherwise.}
%\end{itemize}
%The weight of these quantities in the linear combination is given by a parameter $\alpha$, so that the final similarity measure $P_{ff'}$ reads:
%\begin{equation*}
%    P_{ff'} = \alpha S_{ff'} + (1-\alpha) J_{ff'}.
%\end{equation*}
%The parameter $\alpha$ can be tuned according to the goodness of the prediction on M\&A processes.
%We find that the optimal value of  $\alpha$ depends on the class imbalance of the prediction exercise; a detailed investigation is presented in the Methods section.
    All these measures can be also interpreted as a projection of the company-technology bipartite network onto the company layer; this can be accomplished by using the method of co-occurrences and different normalizations \cite{cimini2022meta}.
    \item Indirect measures. These measures are based on the initial construction of relatedness metrics between technologies and firms and the subsequent evaluation of the relatedness between firms. We employed two different ways to build the relatedness between firms and technologies: one is based on co-occurrences networks, the other on Machine Learning.
    \begin{itemize}
        \item Networks: Following the standard co-occurrence approach \cite{Teece1994understandig}, we projected the bipartite network onto the technology layer using two different normalizations, obtaining two symmetric matrices $B_{tt'}$. We refer to these two normalizations as \textbf{Technology Space} (TS) \cite{Hidalgo2007} and \textbf{Micro-Partial} (MP), based on the work of Teece et al. \cite{Teece1994understandig}. Finally, we compute the prediction scores, called \textit{coherence}, as $\gamma_{ft}^{Y}=\sum_{t'} M_{ft'}^Y B_{tt'}$.
        \item \textbf{Random Forest}: The second approach to the construction of a relatedness measure between firms and technologies is based on the use of a machine learning algorithm. Following \cite{straccamore2022will}, we employ the Random Forest classifier \cite{breiman2001random}, trained on patent data, to predict the technologies that will be patented by firms in the future. The output of this classifier is a score ${RF}^{Y}_{ft}$ which represents the likelihood that the link $M_{ft}^Y$ is $1$. These scores represent an optimal measure of the relatedness \cite{albora2021product,tacchella2021relatedness}, in this case, between a company and a technology.
    \end{itemize}
    Given these different methodologies to assess the relatedness between firms and technologies, we compute the similarity between acquirer and target firms in a M\&A deal as the mean relatedness between the acquirer and all the technologies in the target's portfolio. In the following, we will globally refer to these indirect measures as \textbf{Mean Coherence}:  $\bar{\gamma}_{f}$ and $\Bar{RF}_f$. So the measures we test are Mean Coherence Technology Space (\textbf{MC TS}), Mean Coherence Micro Partial (\textbf{MC MP}) and Mean Coherence Random Forest (\textbf{MC RF}). Due to its definition, $\bar{\gamma}_{f}$ and $\Bar{RF}_f$ are, in general, highly correlated with the diversification of the firm $f$, i.e. the number of technologies to which $f$ is linked, which, in our case, is the acquirer firm in the deal (see Supplementary Information for more details).
    %\textcolor{brown}{MATTEO: sì mettiamole le figure nelle SI, tanto ricordo che le avevamo già fatte. Mettiamo gamma di TS e di RF vs diversification sia per le non normalizzate che non (quindi 4 figure in tutto). In quelle non norm si vedrà la correlazione, nelle altre no}.
    To test if this correlation has some effect on the results of our forecast exercise, we rescaled these two measures between 0 and 1. In this way, we will have rescaled Mean Coherence Technology Space (\textbf{MC TS resc}), rescaled Mean Coherence Micro Partial (\textbf{MC MP resc}) and rescaled Mean Coherence Random Forest (\textbf{MC RF resc}).
\end{itemize}
\begin{itemize}
    \item \textbf{Continuous Company Space} (CCS): These two measures of similarity between firms refer to the construction of the Continuous Projection Space (CPS) \cite{tacchella2021relatedness,straccamore2022will}. CPS represents a way to visualize the similarity between the nodes of one layer of a temporal bipartite network. In this case, we will represent companies in a two-dimensional space. In particular, we build the Continuous Company Space (CCS) starting from two measures of distance between companies, based respectively on Jaffe (we will refer to it as \textbf{CCS Jaffe}) and the Jaffe + Sectors model (we will refer to it as \textbf{CCS J+S}). CPS is instead usually built starting from the prediction scores of a machine learning model \cite{tacchella2021relatedness}; here we build the CPS only using the best performing measures. As we will show in the following, CCS has a minor predictive power than the original distances due to the loss of information in the dimensionality reduction process. Nevertheless, it represents an optimal way to visualize similarities between companies.
\end{itemize}

\subsection*{Visualization on CCS}
\begin{figure}[h!]
\begin{center}
\includegraphics[scale=.9]{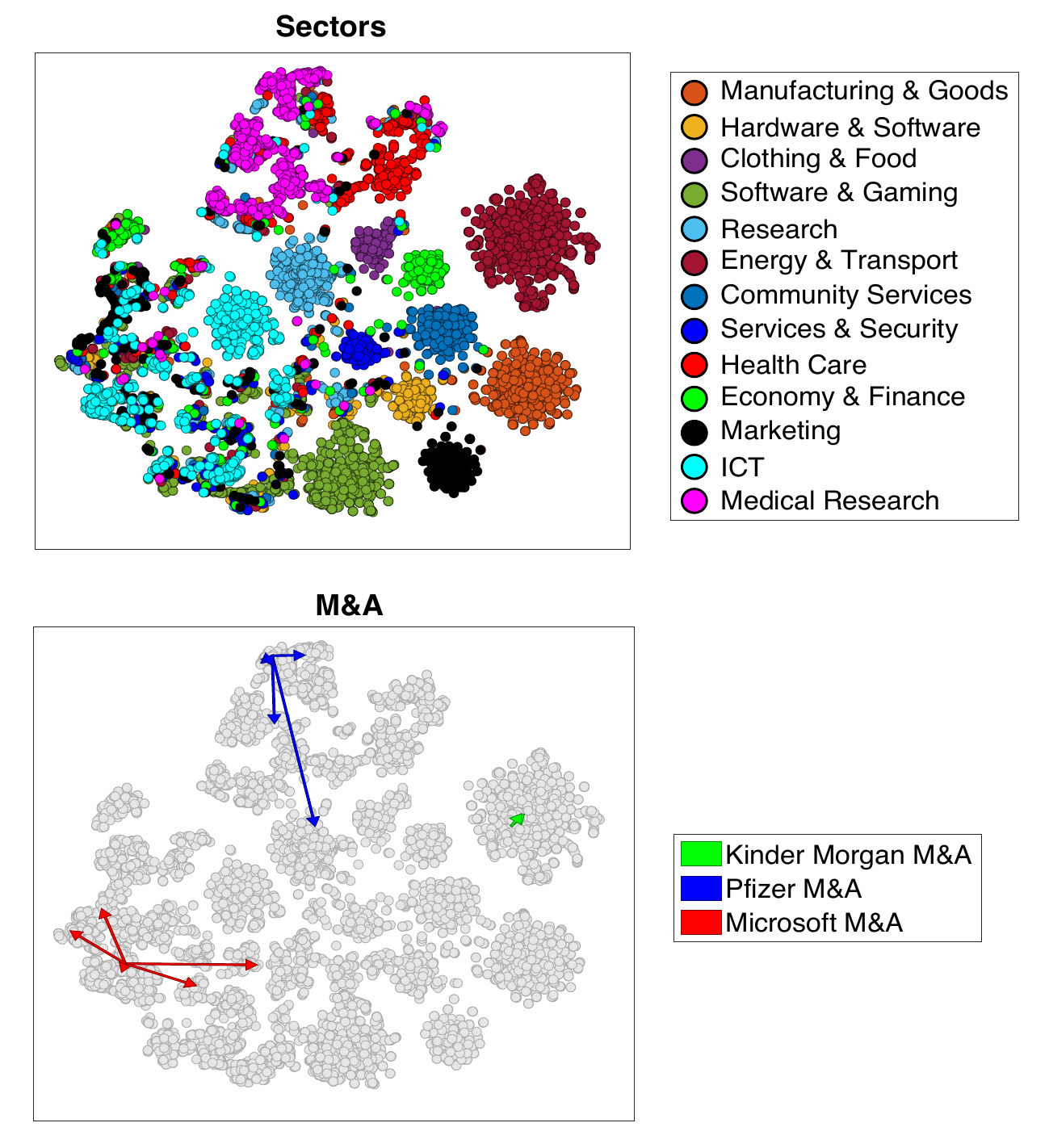}
\caption{\textbf{CCS Jaffe + Sectors ($\pmb{\alpha} = 0.1$).} In the CCS (Continuous Company Space) representations above each point represents a company. On the upper panel, points are colored according to companies' industrial sectors. The effects of the information on the industry are visible from the clustering of firms of the same color. On the lower panel, we represent on the CCS some acquisitions performed by three large companies as arrows from the acquirer firms to the targets. This is an example of how acquisitions are likely to be done locally in this space.}
\label{fig:frecce}
\end{center}
\end{figure}
%We now discuss the Continuous Company Space (CCS) built on the Jaffe + Sectors model. As previously explained, this tool allows simple, 2-Dimensional visualization of the relatedness between companies. 
%In particular, 
In Figure \ref{fig:frecce}, we present the CCS computed starting from the Jaffe + Sectors model with the parameter $\alpha$ fixed to $0.1$. We decided to show the CCS computed also with the indutrial sector information to give an easy visualization of the effects of the last. Here each point represents a company, and the relative distances are a low-dimensional representation of the distances provided by the Jaffe + Sectors model, obtained using the t-SNE algorithm\cite{van2008visualizing}. On the upper panel, we colored the companies according to the respective industrial sector. As expected, since we are using also the information on companies' industrial sectors to build the similarity measure, we find a clear clustering among the firms that belong to the same sector. However, a relative high number of companies end up in a cluster different from the one they should belong to according to the exogenous classification. This is the case, for instance, of many Integrated Control Technology (ICT) companies (light blue), which are spread into different communities on the leftmost side of the plot. These companies have a patenting activity much similar to Marketing companies (black) and Software and Gaming (dark green). A direct consequence of this disposition can be observed in the M\&A behavior. On the lower panel, we show how M\&A processes look like in this space. We draw arrows from 3 big acquirer companies in three different sectors (Pfizer, Microsoft, and Kinder Morgan) to their target firms. As it can be seen, acquisitions are likely done locally in this space, meaning that the proximity measure provided by the CCS can be connected with firms undergoing a merger or acquisition process. Obviously, a company may also decide to perform a deal related to a target that is not close in this space. This is a strategic choice: companies may want to enter into a different area of the CCS and to drastically diversify their technological activity. Also in this case, the CCS provides a map to navigate the innovation space.

\subsection*{Predictions} 
In this section, we present the results of our forecast exercise. We employ all the measures described before to predict which companies will take part in a M\&A process.
We adopted the Best-F1 score to compare the different methods in terms of the goodness of their predictions.
The Best-F1 represents the maximum value that can be obtained by finding the optimal threshold used to compute the F1 score \cite{dice1945measures,van1974foundation} (this metric is discussed in detail in the Methods section). Other performance metrics are discussed in the Supplementary Information. For each acquirer-target pair, we calculate our similarity measures and then we rearrange them to an array of predictions $\textbf{s}$, whose elements and size will be discussed in the following. Then we compare these predictions with the array $\Bar{\textbf{s}}$ whose elements are $1$ or $0$, if the pair is a positive example, i.e. the pair completes an acquisition, or not. To evaluate the goodness of our predictions, we compare $\textbf{s}$ and $\Bar{\textbf{s}}$ computing the Best-F1.\\ % for  and finally we compute the mean over all years.\\
The forecast can be formulated in two different exercises, that lead to different prediction arrays even if the measure is the same:
\begin{itemize}
    \item{\textbf{Pair Prediction}: Given a set of companies, we want to predict which pairs of firms will undergo a M\&A process. This is the prediction task investigated in \cite{FUTAGAMI202122}; here the point of view is of an external observer who compares all the possible pairs.}
    \item{\textbf{Target Prediction}: Given an acquirer company, we want to predict which firm is likely to be its target. Here the point of view is of the acquirer: this firm compares all possible targets with its technological portfolio.}
\end{itemize}
Since our prediction exercise is a classification problem, we need both positive and negative samples. Therefore, for each true acquisition we extract $N$ random examples, with the parameter $N$ controlling the class imbalance of the problem. Because of this the shape of $\textbf{s}$ and $\Bar{\textbf{s}}$ will depend on the class imbalance and will be equal to $N_{\text{true}} \times (1 + N)$. Specifically, in the Pair Prediction case, we extract random pairs of companies, while in the Target Prediction case we extract only the target firms. We repeat this exercise for each class imbalance $20$ times, so extracting different sets of negative samples, and calculating the Best-F1 mean and standard deviation.
\begin{figure}[h!]
\centering
\subfloat[]%[\emph{.}]
   {\includegraphics[width=.90\textwidth]{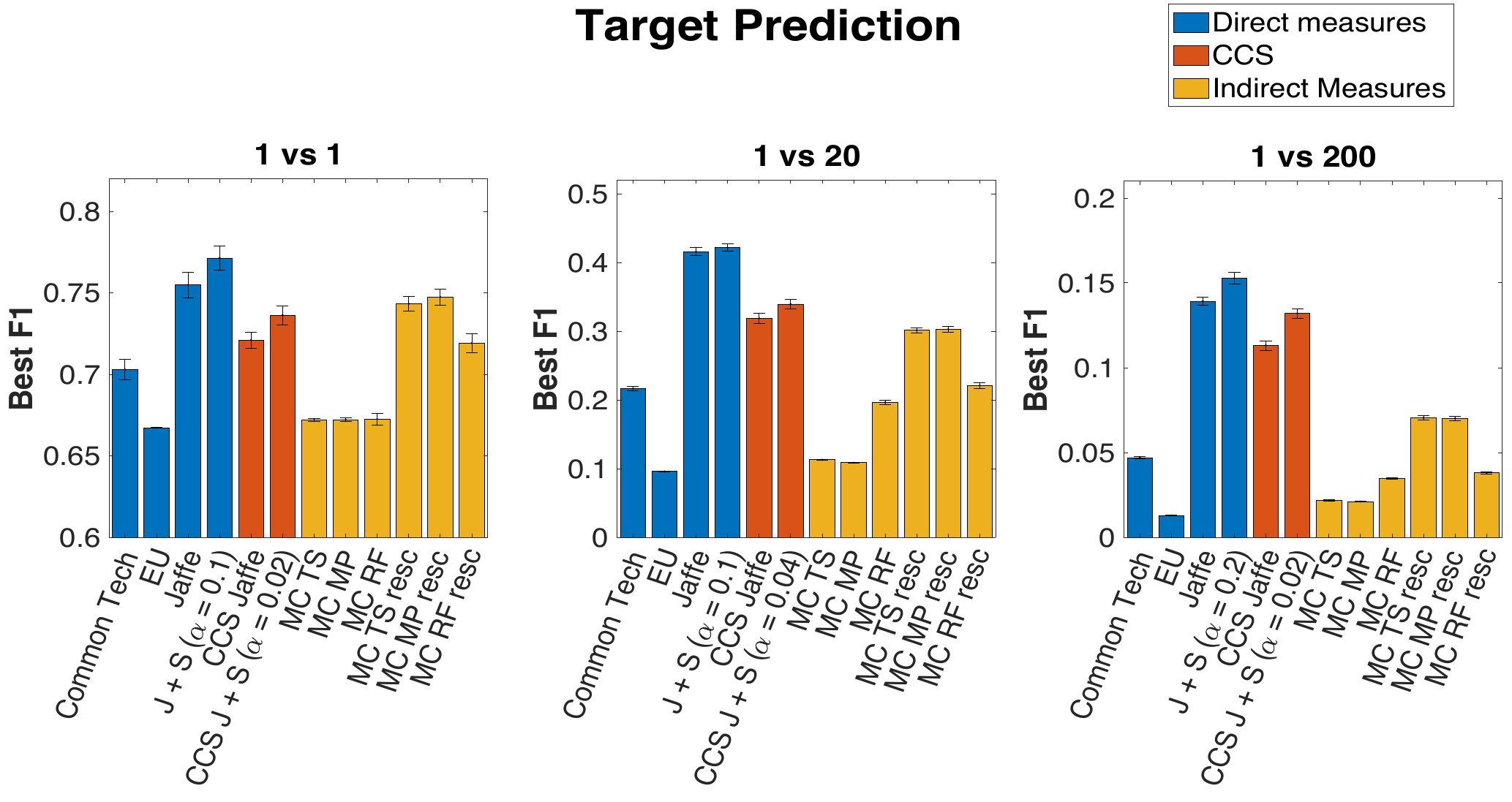}} \\
\subfloat[]%[\emph{.}]
   {\includegraphics[width=.90\textwidth]{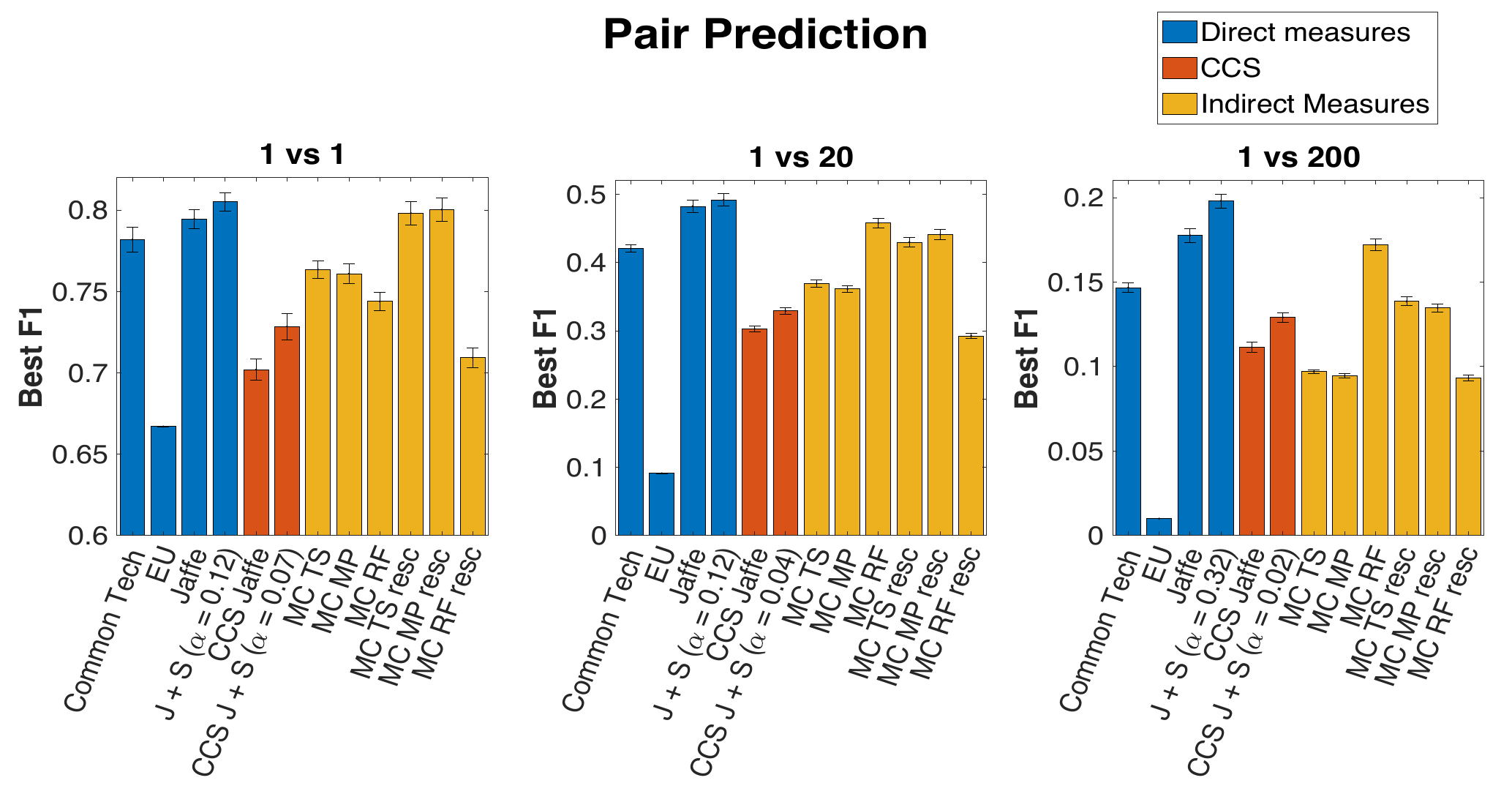}}
\caption{\textbf{Predictions evaluation of Target and Pair forecasts}. Comparison between the prediction performances of the different similarity metrics. We investigated different values of class imbalance, that is for each true M\&A we extract 1, 20, and 200 negative cases (random couples with no deals). In both figures, we use three different colors to distinguish the typologies of metrics. Blue represents direct metrics, yellow indirect ones, and orange the CCS ones. The error bars are computing by repeating the extraction of random Target companies and/or Acquirer companies $20$ times. \textbf{a}: Comparison among the Best F1 scores on Target Prediction (fixed acquirer). \textbf{b}: Comparison among the Best F1 scores on Pair Prediction (both target and acquirer are not fixed). The Jaffe + Sectors metrics outperforms all other approaches.}
\label{fig:TP_PP}
\end{figure}
Notice that, as a robustness check, we evaluated our predictions also using other performance measures than the F1 score. Since the results were fully compatible with the ones shown by the Best F1, we decided to present these in the main text and the others in the Supplementary Information.\\
In Figure \ref{fig:TP_PP}, we show the Best F1 of the prediction tasks for 3 different values of negative samples, and so of class imbalance (1 VS 1, 1 VS 20 and 1 VS 200) both for the Target and Pair prediction.
%From the left we present the results relatives to the measures:
%\textcolor{red}{ok fare un riepilogo ma così è solo un elenco, meglio fare una tabella aggiungendo informazioni. comunque va messo prima, alla fine delle spiegazioni}
%\begin{itemize}
%    \item{\textbf{Common Tech},}
%    \item{\textbf{EU}}
%    \item{\textbf{Jaffe}}
%    \item{\pmb{$J+S (\alpha = \hat \alpha)$}: Jaffe + Sectors model with the paramenter $\alpha$ chosen to maximise the Best F1 at the givem class imbalance,}
%    \item{\textbf{CCS Jaffe}}
%    \item{\textbf{CCS} \pmb{$J+S (\alpha = \hat \alpha)$}}
%    \item{\textbf{MG TS}: Mean $\gamma$ for Technology Space network}
%    \item{\textbf{MG MP}: Mean $\gamma$ for Micro-Partial network}
%    \item{\textbf{MG RF}: Mean $\gamma$ constructed starting from the results of the Random Forest}
%    \item{\textbf{MG TS norm}: MG TS with normalized $\gamma$ }
%    \item{\textbf{MG MP norm}: MG MP with normalized $\gamma$ }
%    \item{\textbf{MG RF norm}: MG RF with normalized $\gamma$ }
%\end{itemize}
%\textcolor{red}{meglio fare figura singola dato che si parla sempre dei confronto tra le due}\\
With each of the three colors, we refer to the three different types of categories of metrics: directed, indirect, and CCS measures.\\

\subsubsection*{General considerations}
One first consideration is that, in general, the Best F1 values of the Pair Prediction task are higher than the ones in the Target Prediction case. This suggests that it is easier to predict which pair of companies is more likely to complete an acquisition than which target will be chosen from a hypothetical acquirer firm. This effect could be a direct consequence of how the tasks are defined. The two tasks differ mainly on the choice of negative samples. In the Target Prediction task we are comparing different possible targets for the same acquirer and some of them can be similar to the acquirer company even if they don't undergo and acquisition process with it.
On the other hand, in the Pair Prediction task, we are comparing the M\&A pairs with randomly assembled pairs of companies that, in most cases, are quite different from each other.
%\textcolor{red}{important, expand: this means that patenting activities are more important? or that the prediction task is easier? why?}
Moreover, as it can be seen from the left panels of the Figures, if we consider a problem with only one negative example for each positive one, the performances of all measures are quite similar to each other. In this case, all Best-F1 values are over 0.65, with the top ones over 0.8 in the Pair Prediction task.
Increasing the class imbalance, differences between metrics become more evident, since the prediction task becomes harder.
In both cases, the highest values of Best F1 are reached by the measure Jaffe + Sectors with the $\alpha$ values chosen a posteriori to maximize the performances at the given class imbalance. As one can see, this measure represents a positive correction to the Jaffe measure, which itself provides the second-best predictive performance among the similarity metrics we studied.
Note that in a high dimensional sparse space as the technologies' one, with 7132 dimensions, it is often recognised that metrics based on cosine similarity or scalar product are a good choice \cite{HAN201239}.\\
The metric with the worst performance turns out to be the inverse euclidean distance \textit{EU}, which is computed in a technology codes' space with 7132 dimensions.
It is known that such a metric loses much of its descriptive power when the dimensionality of the space increases \cite{Aggarwal2002Onthesurprising}. This behavior is usually referred to as the "curse of dimensionality": when the number of dimensions is high, data become sparser and some distance measures (like for example the \textit{$l_k$} ones, with $k>1$) lose informative power. As suggested by the authors of \cite{Aggarwal2002Onthesurprising}, we tried different measures with fractionary $k$, however we did not find any sensible improvement in the results.

\subsubsection*{CCS results}
Regarding the results relative to the CCSs, directly using these metrics for predictions leads to worse results compared to the ones obtained employing Jaffe and Jaffe + Sectors. This is, in general, expected, since the construction of CCSs implies a dimensionality reduction that usually implies a loss of information. Nevertheless, the predictive performance of the CCS remains comparable with the other measures, especially in the 1 vs 1 and in the target prediction exercise.\\
Note that regarding the Jaffe + Sector measure, we performed an optimisation of the Best-F1 with respect to $\alpha$ for the relative CCS metric. 
In this case, the maximum of the Best F1 is less dependent on $\alpha$ and class imbalance than in the original measure. Moreover, best values of $\alpha$ are in most cases around or less than $0.05$: in the CCS only few information on industrial sectors is needed for optimal predictions.
In Figure \ref{fig:TP_PP}, among the other measures, we present the results relative to the CCS Jaffe + Sector with the optimised value of $\alpha$ for each class imbalance.
%In this case, we found out that, while for a class imbalance lower than 1 vs 10 the optimal $\alpha$ is around 0.1 and almost coincide with the Jaffe + Sector optimal $\alpha$, for more extreme class imbalances the best $\alpha$ is 0.
%For this reason, in Figure \ref{fig:TP_PP}, we present the results relative to the CCS Jaffe + Sector measure with the same value of $\alpha$ of the original Jaffe + Sector.
The interested reader can find more details about the optimisation on $\alpha$ in the CCS's case in the Supplementary Information.
An observation regards the difference between Target Prediction and Pair Prediction in the CCS. While in the former the difference between the Best F1 of Jaffe and CCS Jaffe is only 25\%, in the latter the difference rises to nearly 40\%. This can be easily explained saying that in the CCS is more easy make a prediction if we fix the acquirer, i.e. if we use the knowledge about who will do the deal. On the other hand, not doing this would mean considering all possible combinations on the CCS, and this inevitably leads to worse prediction scores. To visualize better, we can imagine that, on the CCS, in the Pair prediction we have to select correctly both where the arrow starts and ends; in Target instead only where it arrives.

\subsubsection*{Rescaling effects}
The performance of indirect measurements (yellow bars in the figure) depends strongly on whether or not we rescale them between 0 and 1. Remember that we use the rescaling to take into account the diversification of companies.
%With diversification we refer to the number of technology sectors in which a companies produces patents. In particular, we refer to the diversification of acquirers because, by construction, these measures refers to those companies.
On average, $\bar{\gamma}_{f}$ and $\bar{RF}_{f}$ are strongly correlated with the diversification of the firm $f$. Here we refer to the diversification of acquirers because, by construction, these measures refer to those companies.\\
With the rescaling, the $\bar{\gamma}_{f}$ predictions clearly improve; the prediction performance of $\bar{RF}_{f}$, instead, increases on the Target Prediction exercise, while decreases in the case of the Pair Prediction. When rescaling, we observe two different effects on the performances that depend on the diversification of the acquirer companies.\\
After the rescaling, $\bar{\gamma}_{f}$ loses most of its correlation with the diversification, while $\bar{RF}_{f}$ maintains it only for high diversified firms (the interested reader can find the relative plots of average \textit{coherence} versus diversification in the Supplementary Information).\\
If the diversification of the acquirers is the same in both positive and negative examples (i.e. real and not real M\&A), rescaling, especially for $\bar{\gamma}_{f}$, the predictive power of the metrics increases.  In fact, if the metrics are not correlated with the diversification we can avoid both part of False Positives and False Negatives. The first ones come from high diversified random extracted acquirers that have, on average, high values of $\gamma$, while the second ones come from low diversified true acquirers that have, on average, low values of $\gamma$.\\
If the diversification of the negative cases' acquirers is lower than the positive cases' ones, the best predictions are made with the metrics without rescaling. In fact, in these cases, the correlation between the metrics and the diversification helps the prediction: not rescaling, we can have more True Positives and True Negatives samples, with respect to the rescaling one. The first ones come from high diversified true acquirers that have, on average, high values of $\bar{\gamma}_{f}$ and $\bar{RF}_{f}$, while the second ones come from low diversified random extracted acquirers that have, on average, low values of $\bar{\gamma}_{f}$ and $\bar{RF}_{f}$. These two effects have two different consequences if we are treating Target or Pair prediction.\\
In the Target Prediction case, acquirers are always the same and so, their diversification doesn't change. We see from Figure \ref{fig:TP_PP}\textbf{a} that rescaling the network-based indirect measures $\bar{\gamma}_f$ leads to better performances, while the $\bar{RF}_f$ one doesn't experience any strong change.
In the Pair Prediction case, we can see in \ref{fig:TP_PP}\textbf{b} a combination of the two described effects on prediction. In fact, the diversification of random companies is lower than the one of acquirers, but this difference is not so wide. As a consequence, for the network-based measures, it is still convenient to rescale the metrics, even if the difference between the Best F1 level in the rescaled and not-rescaled case reduces, in comparison with the Target Prediction task. On the contrary, leaving the $\bar{RF}_f$ measure without rescaling gives better results in predictions. %First, results related to indirect measures in the Pair Prediction case are strongly influenced by the diversification of the randomly extracted companies. \textcolor{red}{ok. figura in SI?}
%In the Target Prediction task we did not observe any significant change in the measure performances, while we did in the Pair Prediction case \textcolor{red}{misterioso. cosa succede?} \textcolor{brown}{MATTEO: Effettivamente qui va spiegato un pochino meglio. Ho capito cosa intendi: nel PP i punteggi gialli si alzano parecchio ed il motivo è che estraendo entrambe le aziende, poiché quelle a bassa div sono di più, è più probabile prendere random una coppia di aziende a bassa diversificazione e quindi automaticamente avranno un'alta similarità. Nel TP, prendendone una sola random, questa sarà a bassa div e quella che fa il MA sarà ad alta div e di conseguenza ecco la bassa similarità. Però quello che è scritto è "nel TP non vediamo cambiamenti significativi nelle metriche, nel PP sì. Se guardiamo le gialle effettivamente è vero ed il motivo è questo che ho anche spiegato io (doppio random --> random anche nelle metriche). Comunque qui farei riassunto della parte che ho spostato nei Methods, nella Section "Effect of rescalation on results".}. In fact, the diversification of companies in the database follows a power-like distribution, where low diversified firms are more numerous. \textcolor{red}{frase sconnessa col resto. che c'entra? spiega. ps quando uno dice power law è perché i valori alti sono più numerosi! qui il punto è il contrario. even if it is power law, the fraction of...}
For the sake of completeness, in the Supplementary Information we also report the analysis done extracting the negative examples among all companies in our dataset, and not extracting them from only the $913$ final companies involved in the M\&A. %\textcolor{brown}{È giusto questo che ho detto? Andrebbero messe altre figure lì. E andrebbe spiegato un pochino meglio.}.

\subsubsection*{Comparison between Jaffe and Common Tech}
We remind the reader that, Jaffe is the cosine angle between two firms in the space of technologies, while Common Tech is the scalar product between them. In other worlds, Common Tech can be interpreted as the module of the magnitude that determines the technological proximity between companies, while Jaffe represents the versor.
%Even if in this case we cannot talk about rescalation in terms in which we have defined it, it is clear that Common Tech is in general correlated with the diversification of the acquirers while Jaffe is not.
For definition, it is clear that Common Tech is correlated with the diversification of the acquirers while Jaffe is not, and this is because the second is the normalized quantity (versor) of the first, i.e. we can see it as a rescaling between 0 and 1. This reflects in the fact that the difference in Best F1 values between the two measures is wider in the Target Prediction than in the Pair Prediction. Jaffe's better performance than Common Tech is telling us that, in the space of technologies, information about the module is not as important as information about the direction.
%\textcolor{red}{vedi sopra}\textcolor{brown}{MATTEO: anche qui ho capito. però il problema è più che altro che uno non ha ancora visto le formule di Jaffe e CT quindi non può ben capire. Forse va detto che jaffe è il versore del prodotto scalare CT: CT è la proiezione di un vettore-azienda su un altro, mentre Jaffe è solo il versore. Questa cosa è interessante effettivamente: Sta dicendo che l'informazione sul modulo non è importante quanto quella della direzione.}

\section*{Discussion}
Mergers and Acquisitions (M\&A) represent a huge market, in which the innovative activity of companies play a major role. In this work we compare mainstream and economic complexity methods to predict M\&A by comparing the technological portfolios of companies as defined by their patenting activity. In particular, we compare different measures of technological similarity between firms to forecast which will be the future deal of M\&A. To the best of our knowledge, this is the first attempt to assess the similarity between firms using Economic Complexity methods. In order to compare the prediction performance of various measures of similarity, we analyze a database consisting of $8737$ firms, of which $913$ are involved in $547$ M\&A deals, and $7132$ technology sectors. We develop a forecasting exercise using the assumption that, on average, a pair of firms will more likely sign a deal if they are similar from a technological point of view. We find that the best performing metric uses the Jaffe cosine similarity between the two technological portfolios combined with the information about the industrial sector. This metric clearly outperforms the standard methodologies usually adopted in economic complexity, that is, networks of co-occurrences. Our results are robust with respect to two different types of forecasting exercises: in the Pair Prediction, we want to forecast the most probable pair of firms; in the Target Prediction , we want to find the best target firm for a specific acquirer. Finally, we discuss the Continuous Company Space (CCS), a visualization tool to represent the proximity between firms in a two-dimensional plane. The CCS can be used to inform strategic M\&A policies; for instance, to plan the attack to a specific market by acquiring a target company specialized therein.

\section*{Materials and methods}
In this Section, we describe in more detail the databases, the proximity measures and the metrics used in the analysis.

\subsection*{Data}
The information used to perform the analysis of the present paper can be obtained from four databases. The two databases AMADEUS and PATSTAT contain the information used for the construction of the bipartite company-technology networks. Zephyr and Crunchbase contain information on the M\&A. The companies' industrial sectors are obtained from the Crunchbase database.

\subsubsection*{Companies}
The information regarding the companies was obtained from the AMADEUS database (\url{https://login.bvdinfo.com/R0/amadeusneo}). This database contains information about over 20M companies located mainly in Europe. AMADEUS is managed by Bureau van Dijk Electronic Publishing (BvD), which specializes in providing financial, administrative, and budget information relating to companies. The BvD includes the same patent identifiers as the European Patent Office and this makes the AMADEUS and PATSTAT databases compatible with each other \cite{pugliese2019coherent}. Although one of the most well-known problems of AMADEUS is that large companies are fully covered while those with fewer than 20 employees are underrepresented \cite{ribeiro2010oecd}, for the purposes of this paper this is not a relevant problem.

\subsubsection*{Technology Codes}
The source of data on technologies and patents is the Worldwide Patent Statistical Database (PATSTAT, \url{https://www.epo.org/searching-for-patents/business/patstat.html}) of the European Patent Office (EPO), which aggregates and organizes data from regional and national patent offices. The most important element of this database is the presence of a standardized code defined within the International Patent Classification (IPC), a hierarchical classification system, internationally recognized, maintained and updated by the World International Patent Organization (WIPO). The codes are organized by levels of increasing aggregation: the lowest level includes over 70,000 groups, while the highest includes only 8 sections. This coding is used to classify each patent from a technological point of view.
For example, the code Axxxxx corresponds to the macro category "Human Needs" and Cxxxxx to the macro category "Chemistry"; considering the following figures we have, for example, with A01xxx the sector "Agriculture; Hunt", and with A43xxx the "Footwear" sector. It is important to note that classes "99" and subclasses "Z" are not considered in this work, as they represent technologies classified in "other classes or subclasses", and therefore are not well defined.
The interested reader can find more details about this data in the work of \cite{pugliese2019coherentSup}.

\subsubsection*{Zephyr and Crunchbase}
Merger \& Acquisition data were acquired from two different databases: Zephyr and Crunchbase. Zephyr (\url{https://www.bvdinfo.com/en-us/our-products/data/greenfield-investment-and-ma/zephyr}) is a commercial database, maintained by the Bureau van Dijk Electronic Publishing (bvd), that contains information on the operations of M\&A, IPO, Private Equity, Venture Capital and related Rumour worldwide. Specifically, in this work, we used the section of the database that concerns companies operating in the biopharmaceutical sector. This section includes information on nearly 4000 deals between 1997 and 2016, and over 3700 companies.
Crunchbase (\url{https://www.crunchbase.com}) is another commercial database, originally created to track start-ups, containing information on public and private companies and related acquisitions, mergers, and investments, globally. Crunchbase's dataset is much larger than the Zephyr one, and contains information on over 100 thousand acquisitions, from 1922, and over a million companies.% Crunchbase data were dumped from the relative website in May 2020.

\subsection*{Data processing}
In this Section, we briefly describe how we combined M\&A data with the information on technological portfolios and the construction of our Industrial sectors classification.

\subsubsection*{M\&A data processing}
To study M\&A processes in relation to the technological portfolios of the companies involved, we linked the Zephyr and Crunchbase datasets to the AMADEUS-Patstat one.
As fully described in \cite{pugliese2019coherent}, the AMADEUS-Patstat data can be seen as a bipartite network where each company, identified by its Bureau van Dijk ID (BVDID), is linked to the technology codes of their patents. The weight of the link between a company and a technology code is proportional to the share of patents, deposited by the company, that contain that technology code. The linking process between the M\&A data to AMADEUS-Patstat is different for each of the two datasets.
In the Zephyr dataset, companies are identified with their BVDID, so it was possible to directly associate them with a technological portfolio.
Starting from a set of 3167 companies involved in M\&A processes, we were able to link 430 of them to the relative technological portfolios. Crunchbase's data are not labeled by the BVDID, so we had to match the names of Crunchbase's companies to the AMADEUS ones in order to find for each firm the relative BVDID.
For a better match, companies' names underwent a "cleaning" process to remove symbols, punctuation, and companies acronyms. The full process of names' cleaning and matching is described in the Supplementary Information. After the cleaning process, we ended up with $28137$ companies with a BVDID associated.  From this set, we linked $12017$ companies to the relative technological portfolio. Sometimes, due to the cleaning of names, companies resulted associated with multiple BVDID and thus multiple portfolios. These are occurrences in which, for example, a multinational company has multiple BVDID associated with the various national subsidiaries. In such cases, we merged all the technological portfolios associated with that company. Finally, for each year, in both Zephyr and Crunchbase cases, we kept only the M\&A that happened between 2002 and 2012, whose acquirer and target companies deposited at least one patent from 2000 to that year.
With this constraint, we managed to build a data set of $1279$ M\&A (126 from Zephyr and 1153 from Crunchbase), that involves $1974$ companies ($145$ from Zephyr and $1858$ from Crunchbase, with $29$ present in both).

\subsubsection*{Sectors classification}
Crunchbase companies are organized, concerning their industrial sector, in two levels of aggregation: the lower level counts $744$ \textit{categories}, while the upper counts $43$ \textit{category groups}.
This classification is not directly linked to the official ones (NACE, NAICS, SIC, etc.) but was built independently by Crunchbase.
In this classification, each company is assigned several category groups and thus many categories. 
Starting from this classification we built another level of aggregation consisting of $13$ sectors. In this way, we managed to assign one univocal industrial sector to $8069$ firms, nearly 70\% of the Crunchbase companies that we had previously linked to their technological portfolio.
To have at least a sector linked to each company and a smaller number of sectors is fundamental for the construction of our modified version of CCS and its visualization.
Further details on how our classification was built can be found in the Supplementary Information.\\
The main results presented in this paper were obtained working on a subset of the M\&A data set that includes only companies with univocal sectors assigned within our classification. This subset counts $8737$ companies and $547$ M\&A that involve $913$ companies of the total subset.

\subsection*{Data processing}
Our final data can be used to construct $13$ bipartite networks between companies and technology codes, one for each year from $2000$ to $2012$. We can represent these networks, for each year $y$, as a matrix with elements $M_{ft}^y$. Each matrix element represents the weight of the link between the firm $f$ and the technology code $t$, in the year $y$; this is equal to the (possibly fractional) number of patents filed by $f$ belonging to the technology $t$.
Under the hypothesis that a patent filed in a certain year is representative of the firm's capabilities also in the following years, for the construction of our relatedness measures we consider a summed version of the matrix $M_{ft}^y$ over the years.
We define $M_{ft}^Y$ as the sum of all $M_{ft}^y$ from 2000 to the year $Y$.
From now, we drop the apex $Y$ for simplicity, keeping in mind that all measures can be defined for each year.

\subsection*{Direct measures}

\subsubsection*{Common Tech and Jaffe}
The Common Tech and Jaffe metrics consist of a direct projection onto the companies' layer to measure similarity between each companies pair. We calculate these two quantities by computing the equations:
\begin{equation*}
    \text{Common Tech}_{ff'} = \displaystyle \sum_t M_{ft} M_{f't},
\end{equation*}
\begin{equation*}
    \text{Jaffe}_{ff'}\ = \frac{\sum_t M_{ft} M_{f't}}{\sqrt{\sum_{t} M_{ft}^2}\sqrt{\sum_{t} M_{f't}^2}}.
\end{equation*}
where $M_{ft}$ is the adjacency matrix that link firms to technologies. The element of these matrices represents how much a technology $t$ is present in the patenting activity of firm $f$.
The former is a simple scalar product between the technological portfolio of the two firms. This is correlated with the diversification of both $f$ and $f'$.
The latter is a cosine similarity between the two portfolios, introduced in this context by Jaffe \cite{Jaffe1986technological}. It is bounded between 0 and 1.\\
These measures represent a projection of the bipartite network onto the firms' layer, so they can be interpreted as the weight of a link that connects the companies $f$ and $f'$ in a monopartite network of firms.

\subsubsection*{Euclidean Distance}
To build a relatedness measure between companies based on the euclidean distance we start from the matrix $M_{ft}$.
Each row of this matrix can be seen as the list of coordinates of each company in the space of technology codes.
The relatedness measures $EU$ is just the inverse of the euclidean distance between companies in this space:
\begin{equation*}
    EU_{ff'}\ = \left( \sum_t \left( M_{ft}-M_{f't} \right)^2 \right)^{-\frac{1}{2}} 
\end{equation*}

\subsubsection*{Jaffe + Sectors and best $\alpha$ identification}
The Jaffe + Sectors scores are computing by considering not only the technological affinity between companies but also the industrial sector. This leaves a degree of freedom (the relative weight) which we optimize as described in the following. The formula is:
\begin{equation}
    P_{ff'} = \alpha S_{ff'} + (1-\alpha) J_{ff'},
    \label{eq:JSSS}
\end{equation}
where $J_{ff'}$ is the Jaffe measure between firms' technological portfolios and $S_{ff'}$ is $1$ if both firms belong to the same sector and $0$ otherwise.
The weight of these two pieces of information is controlled by the parameter $\alpha$, bounded between $0$ and $1$. The higher $\alpha$, the greater the importance of the sectors' similarity on the measures.
To understand the behavior of this measure it is useful to consider that the M\&A pairs are distributed in 4 sets according to the respective sectors and the relative distance: 
\begin{itemize}
    \item{\textbf{$S1J1$}: M\&A with $S_{ff'} = 1$ and $J_{ff'} \neq 0$,}
    \item{\textbf{$S1J0$}: M\&A with $S_{ff'} = 1$ and $J_{ff'} = 0$,}
    \item{\textbf{$S0J1$}: M\&A with $S_{ff'} = 0$ and $J_{ff'} \neq 0$,}
    \item{\textbf{$S0J0$}: M\&A with $S_{ff'} = 0$ and $J_{ff'} = 0$,}
\end{itemize}
Due to the fact that $S_{ff'}$ can be only 0 or 1, while $J_{ff'} \in [0,1]$, for $\alpha > 0.5$ all the Best F1 results are independent of $\alpha$ and equal to the one at $\alpha = 0.5$.
In fact, for $\alpha > 0.5$ the elements in the four sets are bounded within their set: in $S1J1$, $P_{ff'} > \alpha$, in $S1J0$, $P_{ff'} = \alpha$, in $S0J1$, $0 < P_{ff'} < (1-\alpha)$ and in $S0J0$, $P_{ff'} = 0$.
Certainly, items in $S1J1$ are classified as positives, while items in $S0J0$ are classified as negatives so the threshold that defines the Best F1 must lie among the elements of $S0J1$ and $S1J0$.
\begin{figure}[h!]
\begin{center}
\includegraphics[scale=0.6]{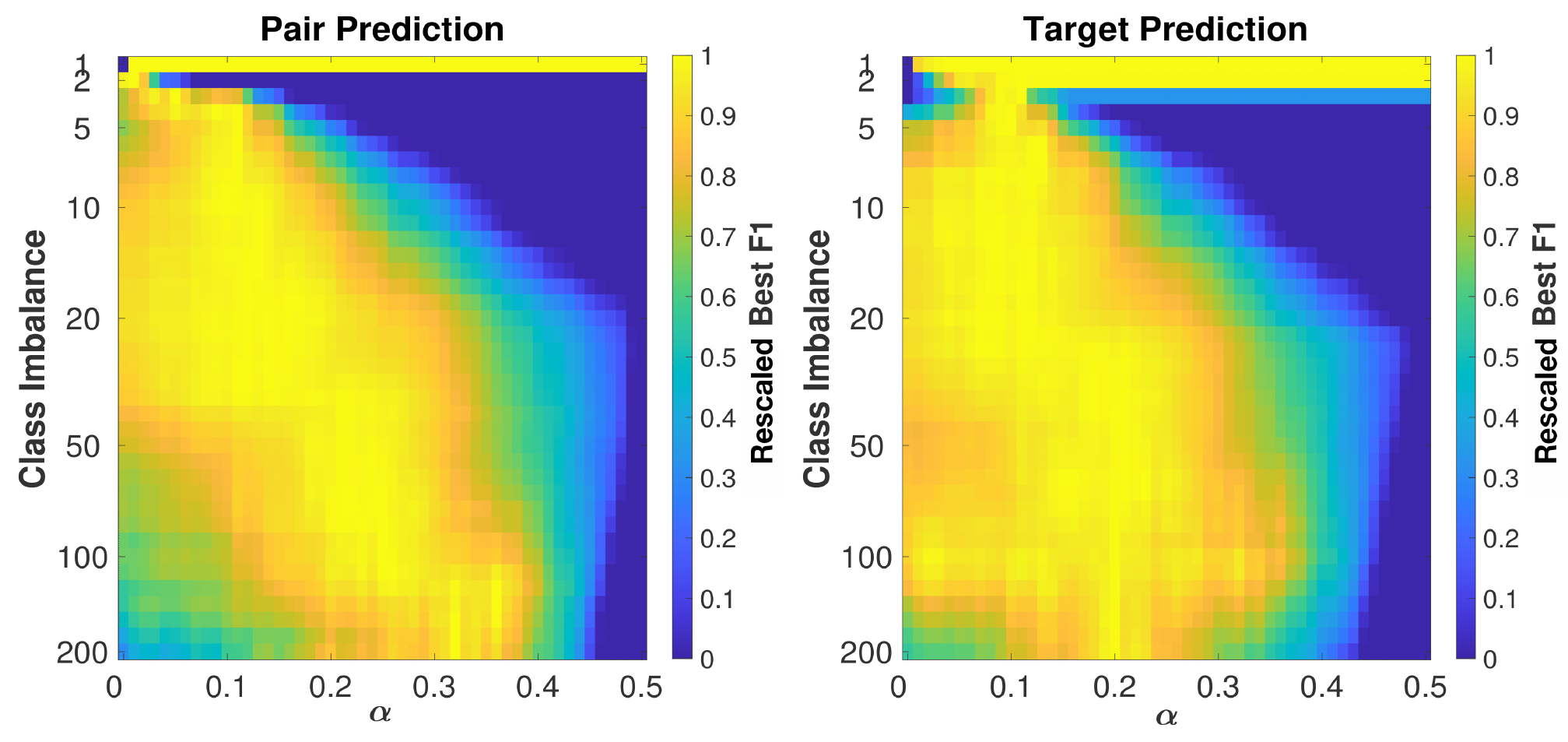}
\caption{\textbf{Dependence of maximum Best F1 on $\alpha$ and the class imbalance in the Jaffe + Sectors measure}. For each value of class imbalance, we rescale the Best F1 between $0$ and $1$ to better visualize the maximum as a function of $\alpha$. The maximum F1 moves towards higher $\alpha$ values when the class imbalance increases. This suggests that when choosing a M\&A pair among a large pool of options, the industrial sector play a more important role.}
\label{fig:3d_bf1}
\end{center}
\end{figure}
Finally, if we sort these elements in descending order, their order does not depend on $\alpha$, thus neither the threshold nor the Best F1 does.
For this reason, we examine the behavior of the measure for $\alpha \leq 0.5$.
Another factor that influences the performance of the measure is the class imbalance, which is defined by the parameter $N$, namely the number of negative examples per positive example.
\begin{figure}[h!]
\begin{center}
\includegraphics[scale=.60]{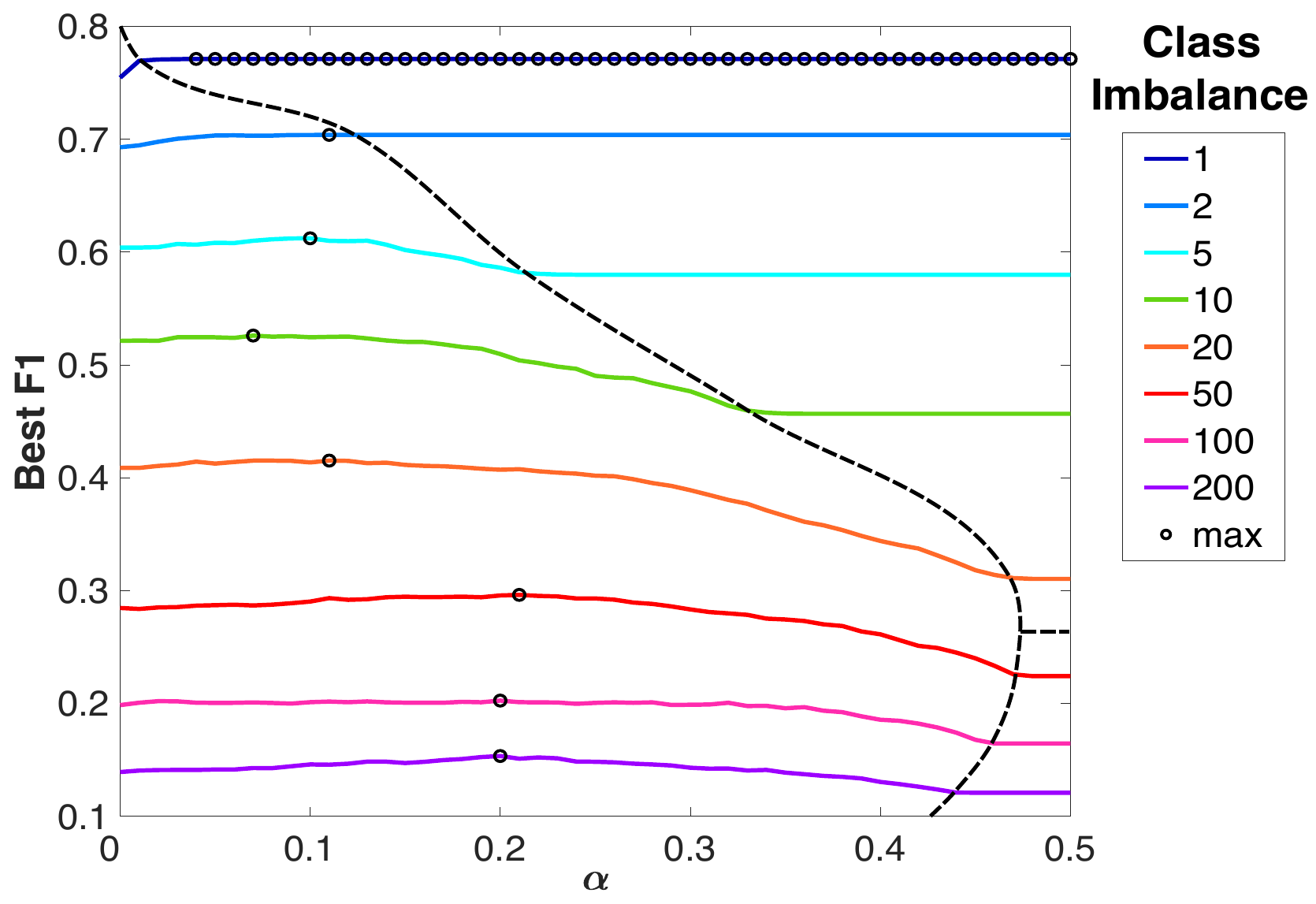}
\caption{\textbf{Behavior of Best F1 versus $\alpha$ for various class imbalance values in the Jaffe + sectors measure.} Due to the correlation between Best F1 and class imbalance the various curves never intersect. In this plot, three phases divided by dotted lines can be spotted: a Low $\alpha$ phase, at the bottom-left, where there exists an $\alpha$-dependent maximum for the Best F1, a High $\alpha$ - Low-class imbalance phase, at the top-right, and the High $\alpha$ - High-class imbalance phase at the bottom-right, where the Best F1 is independent of $\alpha$.}
\label{fig:max_bf1}
\end{center}
\end{figure}
In Figure \ref{fig:3d_bf1} we present the behavior of the Best F1 as a function of $\alpha$ and the class imbalance $N$, both for the Pair Prediction and the Target Prediction. To average out the possible fluctuations coming from the random extraction of negative examples, each point in the Figures reports the average Best F1 over 20 realizations of the prediction exercise.
Because the Best F1 is highly correlated with the class imbalance \cite{jeni2013facing}, for each value of $N$ we rescaled the Best F1 between 0 and 1. In this way, it is possible to better spot the relative maximum of Best F1 as a function of $\alpha$, for each value of class imbalance.\\
From both the figures it emerges that if we increase the class imbalance, the maximum of Best F1 moves \cite{jeni2013facing} towards a higher value of $\alpha$, especially in the Pair Prediction case. This suggests that when choosing a M\&A pair among a large pool of options, the similarity between the companies from the point of view of the industrial sector plays a more important role.
To deepen the understanding of the model, we also considered the absolute value of Best F1, as it is shown in Figure \ref{fig:max_bf1}. In this plot, we show the Best F1 vs. $\alpha$ curves relative to the Target Prediction case for some values of class imbalance; the number of negative samples correspond to lines of different colours. Since the Best F1 is highly correlated with the class imbalance the curves for different values of $N$ never touch each other and can be represented in a single figure.
In this case, the figures relative to Pair Prediction and Target Prediction were almost identical, so we decided to show only one of the two.\\
From Figure \ref{fig:max_bf1}, it is possible to identify 3 phases, divided by dotted lines, for the Best F1:
\begin{itemize}
    \item \textbf{Low \pmb{$\alpha$} phase}. Located on the left of the figure; the Best F1 is $\alpha$-dependent. First it increases, it reaches a maximum and then decreases.
    \item \textbf{High \pmb{$\alpha$} - Low class imbalance phase}. Located at the top-right of the figure; it presents a fixed value of Best F1 independent from alpha.
    \item \textbf{High \pmb{$\alpha$} - High class imbalance phase}. Located at the bottom-right of the figure; it presents a fixed value of Best F1 independent of alpha.
\end{itemize}
To understand this behavior we need to define some other sets in which the M\&A examples can be divided with respect to the measure $P_{ff'}$. Specifically these are three subsets of $S0J1$, so they have $S_{ff'} = 0$ and differentiate according to the $J_{ff'}$ value:
\begin{itemize}
    \item{\pmb{$J_+$}: M\&A with $(1-\alpha)J_{ff'} > \alpha$,}
    \item{\pmb{$J_-$}: M\&A with $(1-\alpha)J_{ff'} < \alpha$,}
    \item{\pmb{$J_\alpha$}: M\&A with $(1-\alpha)J_{ff'} = \alpha$. In this subset $P_{tt'} = \alpha$ as in the $S1J0$ set, so in the subsequent analysis, it will be considered jointly with this last one.}
\end{itemize}
Now, if we consider the $P_{ff'}$ values for each set in ascending order we have: $S0J0$ (i.e. M\&A with $S_{ff'} = 0$ and $J_{ff'} = 0$ in Eq. \ref{eq:JSSS}) with $P_{ff'} = 0$, $J_-$ with $0< P_{ff'} < \alpha$; $S1J0$ (i.e. M\&A with $S_{ff'} = 1$ and $J_{ff'} = 0$ in Eq. \ref{eq:JSSS}) with $P_{ff'} = \alpha$; $J_+$ and $S1J1$ (i.e. M\&A with $S_{ff'} = 1$ and $J_{ff'} \neq 0$ in Eq. \ref{eq:JSSS}) with $P_{ff'} > \alpha$.
The observations on the behavior of the Best F1 are similar in style to the ones made before for $\alpha > 0.5$.
In the upper-right side of Figure \ref{fig:max_bf1}, in the High $\alpha$ - Low-class imbalance phase, the Best F1 is independent of $\alpha$ because the relative threshold is always less than $\alpha$. In this case, the threshold lies between the elements of $J_-$, whose order (in terms of the magnitude of $P_{ff'}$) is independent of $\alpha$.
On the contrary, in the Low $\alpha$ phase, the threshold is always above $\alpha$. This means that it lies among the elements of $J_+$ and $S1J1$.
If we order these elements according to their $P_{ff'}$ value, the two sets are, in general, mixed and the order of elements depends on $\alpha$. So the Best F1 is $\alpha$ dependent.
This remains true until both $\alpha$ and the class imbalance become too large. In this phase, the High $\alpha$ - High-class imbalance one, the $J_+$ set becomes negligible to the $S1J1$, and the threshold always falls between the elements of $S1J1$. Therefore, also in this phase, the Best F1 is independent of $\alpha$.\\
From a practical point of view, this means that, when looking for the best $\alpha$ to optimise the prediction performances at fixed class imbalance, one should always search in the Low $\alpha$ phase. In fact, only in that phase performance metrics like Best F1 are $\alpha$ dependent and can reach a maximum.

\subsection*{Indirect measures}
The idea of these measures is to compute firstly a measure of similarity between technology or of relatedness between companies and technologies, and secondly a measure of similarity between companies. Even if indirect, these approaches are discussed here because they use well-known tools of both mainstream and economic complexity literature.
\subsubsection*{Network-based approaches}
The network-based approach to the construction of a relatedness measure between companies starts from the observation that our data constitute a bipartite network between companies and technology codes described by $\textbf{M}$ matrices. The described bipartite network can be naturally projected onto both the company layer and the technology layer using the co-occurrences method. This gives us the possibility to construct different measures of relatedness.\\
The construction of a relatedness measure based on the projection of the bipartite network onto the technologies' layer follows the work of Pugliese et al. \cite{pugliese2019coherent}.
Firstly, we build a measure of relatedness between technologies, which is the weight of the link between technology codes in their monopartite network. We employ two kinds of normalizations for this measure.\\
The first is the one introduced in \cite{HidalgoK2007} and normalizes the co-occurrences of two technology codes in the same portfolio by the maximum of their ubiquity. We refer to this measure as \textit{Technology Space}.
\begin{equation*}
B_{tt'}^{TS} = \frac{1}{max(u_t,u_{t'})}\sum_f M_{ft}M_{ft'}.
\end{equation*}
where  $u_t=\sum_{f} M_{ft}$ is called ubiquity.
Normalizing by the maximum of the ubiquities allows us to weigh less the co-occurrences between highly ubiquitous technologies.
Moreover, this normalization avoids undesirable effects caused by technologies patented by a single firm.
In fact, if the technology $t$ is patented only by the firm $f$, for every other technology code $t'$, $\sum_{f'} M_{f't} M_{f't'} = 1$.\\
The other type of normalization we make use of in this work was firstly introduced by Teece et al. \cite{Teece1994understandig} in the context of firms' diversification; it has been later employed in several other studies on technological diversification \cite{Piscitiello2000relatedness,Breschi2003knowledge,Bottazzi2010measuring}. We refer to this measure as \textit{Micro-Partial} because the null model hardly constraints one layer and randomizes everything else \cite{cimini2022meta}.
To calculate the Micro-Partial measure we start from a matrix whose elements count the co-occurrences between technology codes within companies' portfolios $C_{tt'}$.
\begin{equation*}
C_{tt'} =  \sum_f M_{ft}M_{ft'}.
\end{equation*}
The $C_{tt'}$ matrix is then normalized with respect to a null model in which technologies are randomly assigned to companies' technological portfolios, keeping their ubiquity fixed.
If we call $u_t$ the ubiquity of the technology $t$ and N the total number of companies, the random variable $x_{tt'}$, the number of companies innovating in technologies $t$ and $t'$ in the random case, follows a hypergeometric distribution with average and variance:
\begin{align*}
 \mu_{tt'} &= \frac{u_t u_{t'}}{N}, & \sigma^2_{tt'} = \mu_{tt'}  \frac{(N-u_t)(N-u_{t'})}{N(N-1)}.
\end{align*}
We therefore define:
\begin{equation*}
B_{tt'}^{MP} =   \frac{C_{tt'}- \mu_{tt'}}{ \sigma_{tt'}},
\end{equation*}
which, in a similar fashion to a t-Student variable, measures the number of $\sigma_{tt'}$ the observed value of $C_{tt'}$ deviates from $\mu_{tt'}$.
In other words, we are comparing the weight of the links in $C_{tt'}$ with the average values generated by a partial Microcanonical null model, in which only one of the two layers, the technologies' one, is fixed. % definisci gamma e mean gamma
Given these two measures of similarity between technology codes, we define the \textit{coherence}\cite{pugliese2019coherent} between a firm $f$ and a technology $t$ as:
\begin{equation*}
\gamma_{ft} =   \sum_{t'} M_{ft'} B_{tt'}.
\end{equation*}
For our scopes, we need a similarity measure between companies. Therefore, to study the relatedness between the two firms involved in a M\&A, we look at the mean coherence (Mean $\gamma$) between the acquirer firm $f$ and technologies of the target company $f'$
\begin{equation*}
\bar{\gamma}_{f} =   \frac{1}{d_{f'}} \sum_{t \in f'} \gamma_{ft},
\end{equation*}
where $d_{f'}$ is the diversification of the target firm, i.e. the number of technologies in its portfolio.\\
Finally, as we mentioned in the Results section, the coherence $\gamma_{ft}$ is correlated with the number of technologies $f$ is linked to, i.e. the diversification $d_{f'}$ of the acquirer. To test if this correlation has some effect on the results of our forecast exercise, we compute a rescaled version of the coherence measure. We call $\hat{\gamma}_{f}$ the vector of all the values of coherence between the firm $f$ and all the technologies (i.e. the $f$-th row of the matrix with elements $\gamma_{ft}$) and we rescale each of these vectors between 0 and 1:
\begin{equation*}
    \hat{\gamma}'_{f} = \frac{\hat{\gamma}_{f} - \min_t(\hat{\gamma}_{f})}{\max_t(\hat{\gamma}_{f}) - \min_t(\hat{\gamma}_{f})}.
\end{equation*}
In this way, the coherence metric $\gamma_{f}'$ is less dependent on the diversification of the acquirer firm.

\subsubsection*{Random Forest}
To compute the relatedness between companies and technologies, following \cite{straccamore2022will}, we use the Random Forest algorithm\cite{breiman2001random}. In particular, we train one model for each target technology.\\
%The Random Forest is a tree-based algorithm that manages to correct some of the relative problems of the Decision Tree, i.e. high variance and high overfitting. As for the high variance, this indicates that if we take a tree and train it using a training set, the result of the prediction on a subsequent test will be very different from the same decision tree that has different training and a test ( but still belonging to the same database). Random Forest solves this problem by averaging the results of multiple decision trees. The overfitting problem is because a simple tree, after even a few splits within it, adapts too much to the training data. The Random Forest solves this problem using a process called Bootstrap, that is, in each division of each branch only a part of the complete input set is taken into consideration.\\
%The use of Random Forest in this study relates to the work of Straccamore et. all \cite{straccamore2022will}.
In general, for supervised machine learning algorithms we have to quantify three quantities:
\begin{itemize}
    \item The matrix of samples $\textbf{X}$. The rows represent the different samples (the companies) that we have to classify correctly. Each element of each row represents a feature (a technology). In our case, $\textbf{X}$ is the matrix obtained by concatenating vertically each matrix $\textbf{M}$ with $y\ \in\ [2000,2010]$ considering the first 10K companies with higher diversification (10K HD). In \cite{straccamore2022will}, authors show how to train RF with the higher diversification firms can increase the forecast results. Also to be consistent with this work, we decide to use only in this case the single $\textbf{M}^y$ and not the $\textbf{M}^Y$s obtained by adding the $\textbf{M}^y$s together.
    \item The vector of the classes $\textbf{y}$: in a generic classification problem it is a vector which the class of the sample is associated to. In our work, we are treating it as a binary classification problem so each element of $\textbf{y} = [0,1]$. So for each model $\textbf{y}$ is a column of the matrix $\textbf{M}^y$ shifted by two years ( with respect to the one used in the training, so $y \in [2002,2012]$) and that we have binarized by setting the elements equal to $1$ if that technology will be made by the firm after two years, and $0$ otherwise i.e. we put $1$ if the element of matrix is different from $0$, and if it is equal to $0$, we leave it as it is.
\end{itemize}
We use these two elements to train our Random Forest, i.e. to learn how the features of the samples (that is, the technologies of the companies) are associated to patenting/not patenting in the target technology.
\begin{itemize}
    \item The matrix of samples $\textbf{X}_{\text{test}}$: after the training, we have to test the performance of our algorithm and this is done using samples that are never seen by it in the training process. $\textbf{X}_{\text{test}}$ is the matrix in which we find these samples. In our case, we use as $\textbf{X}_{\text{test}}$ the matrix obtained by concatenating vertically each matrix with $y\ \in\ [2000,2010]$ considering the companies about which we have either sector or M\&A information. The companies present in $\textbf{X}_{\text{test}}$ and which would have been among the 10K HD used in the training have been removed from $\textbf{X}$ to avoid overfitting problems.
\end{itemize}

Finally, the output of the Random Forest is a matrix that contains the relatedness $RF_{ft}$ between companies and technology codes. We can interpret this measure as another form of \textit{coherence} between firms and technologies and use it in the same way.
Therefore, to study the similarity between two firms involved in a M\&A, we look at the mean values of $RF_{ft}$, i.e. $\bar{RF}_f$ where $f$ is the acquirer firm, and the average computed on all the technologies of the target firm $f'$
\begin{equation*}
\bar{RF}_{f} =   \frac{1}{d_{f'}} \sum_{t \in f'} RF_{ft}.
\end{equation*}
As for the coherence, also in this case the $RF_{ft}$ is correlated with the number of technologies $f$ is linked to, and also in this case we rescale it by
\begin{equation*}
    \hat{RF}'_{f} = \frac{\hat{RF}_{f} - \min_t(\hat{RF}_{f})}{\max_t(\hat{RF}_{f}) - \min_t(\hat{RF}_{f})},
\end{equation*}
where $\hat{RF}_{f}$ is the $f$-th row of the matrix with elements $RF_{ft}$.

\subsection*{Continuous Company Space (CCS)}

To obtain a two-dimensional representation of the proximity between companies, and therefore to obtain a greater interpretability of the results, we introduce the Continuous Company Space (CCS). Tacchella et. all have proposed in \cite{tacchella2021relatedness} the Continuous Projection Space applying it to the exported products, and Straccamore et. al \cite{straccamore2022will} have reformulated this concept applying it to technologies.\\
The construction of our CCS starts from a matrix of distances $D_{ff'}$ between companies.
In particular, we consider the subset of $8079$ companies to which it was possible to assign a sector in our classification. We use two distance matrices.
The first is derived from the Jaffe measure between companies. Since Jaffe's is a similarity measure, to obtain a distance we consider:
\begin{equation*}
    D_{ff'}^J = 1 - J_{ff'}.
\end{equation*}
The other distance measure is derived from the Jaffe + Sectors approach. In this case we consider:
\begin{equation*}
    D_{ff'}^{JS} = \alpha (1-S_{ff'}) + (1-\alpha)D_{ff'}^J.
\end{equation*}
where $J_{ff'}$ is the Jaffe measure between firms' technological portfolios and $S_{ff'}$ is $1$ if both firms belong to the same sector and $0$ otherwise.\\
The columns of the distance matrix can be seen as the coordinates in a high-dimensional space for each company, with a dimension equal to $8079$. Because it is impossible to visualize these coordinates in this such a high dimensional space, we project it in a $2D$ space we call CCS.
Following \cite{tacchella2021relatedness}, this operation consists of two steps. First, we reduce the number of dimensions from $8737$ to $150$ using a Variational - Autoencoder Neural Network \cite{kingma2013auto}. Then, we reduce from $150$ to $2$ dimension using the t-SNE algorithm \cite{van2008visualizing}. Within the $2D$ representation, the similarity between companies is simply given by the relative euclidean distance.

\subsection*{Prediction performance metric: Best-F1}
To better make a comparison between all the algorithms and techniques used in this work, we use as a performance metric the Best-F1 \cite{cruz2016tackling,albora2021product,albora2022machine,tacchella2021relatedness}.
The F1 score is defined as:
\begin{equation}
    F1 = 2\left(\frac{1}{\text{precision}(\tau)}+\frac{1}{\text{recall}(\tau)}\right)^{-1}
\end{equation}
i.e. is the harmonic mean between precision $= \frac{TP(\tau)}{TP(\tau)+FP(\tau)}$ and recall $= \frac{TP(\tau)}{TP(\tau)+FN(\tau)}$, where $TP$ represents the number of True Positives (i.e. the elements equal to 1 that are correctly predicted) and, analogously, $FP$ are the False Positives and $FN$ the False Negatives. These quantities depend on the scores’ binarization threshold $\tau$, that is, the number above which the prediction score is associated with a predicted 1 (if the score is lower than $\tau$, the measure predicts a zero). The Best-F1 is the metric associated with the value of $\tau$ that maximizes the F1 a posteriori. Note that the highest possible value of Best-F1 is 1, which indicates that both precision and recall are equal to 1, and the lowest possible value is 0 if one of the precision or recall is zero.

%\section*{Conclusion}

\section*{Data availability statement}
The data that support the findings of this study are available upon reasonable request from the authors. The original datasets are available upon subscription.

\section*{Acknowledgments}
The authors thank Louis Barlascini for his preliminary investigations, Arianna Martinelli for providing the Zephyr data, Lorenzo Napolitano for providing the bipartite company-technology data, and Luciano Pietronero for scientific discussions.

\bibliography{sample}

\end{document}

\title{\huge{Supplementary Information}\\\Large Prediction and visualization of Mergers and Acquisitions using
Economic Complexity}

\author{Lorenzo Arsini, Matteo Straccamore and Andrea Zaccaria}

\date{}
\maketitle

\newpage

\tableofcontents
\newpage

\newpage

\section{Pair Prediction on the full database}
As stated in the Result section, performances of some measures employed in this work, in the Pair Prediction task, can depend on the diversification of companies that are chosen for the negative examples.
\begin{figure}[h!]
\begin{center}
\includegraphics[scale=.50]{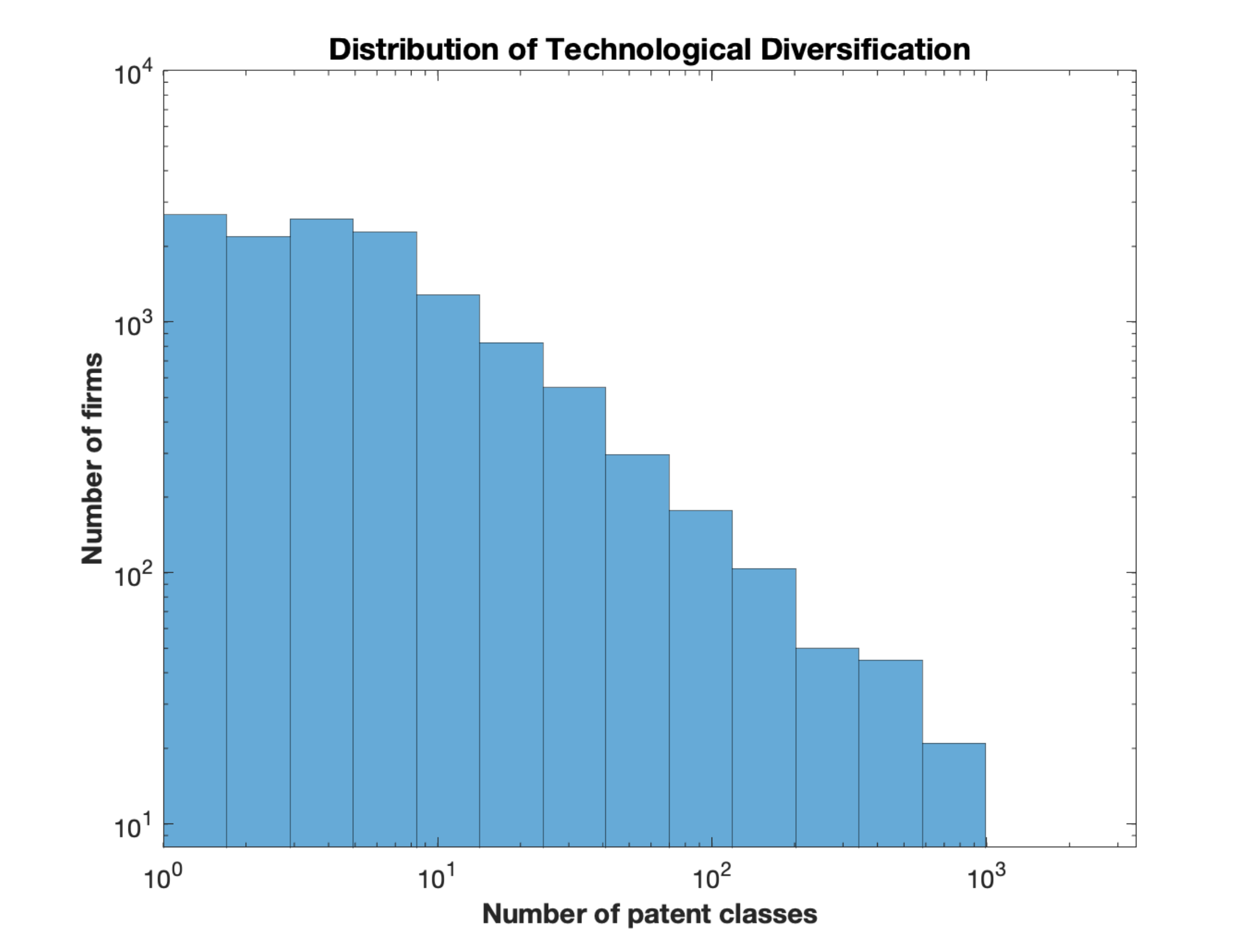}
\caption{\textbf{Distribution of technological diversification.} The diversification of companies in our database follows a power-like distribution, where low diversified firms are prevalent, and this implies that, if the
negative examples companies are drawn from the entire database, using metrics which is correlated with diversification leads to better predictions.}
\label{fig:DistPat}
\end{center}
\end{figure}
\begin{figure}[h!]
\begin{center}
\includegraphics[scale=.8]{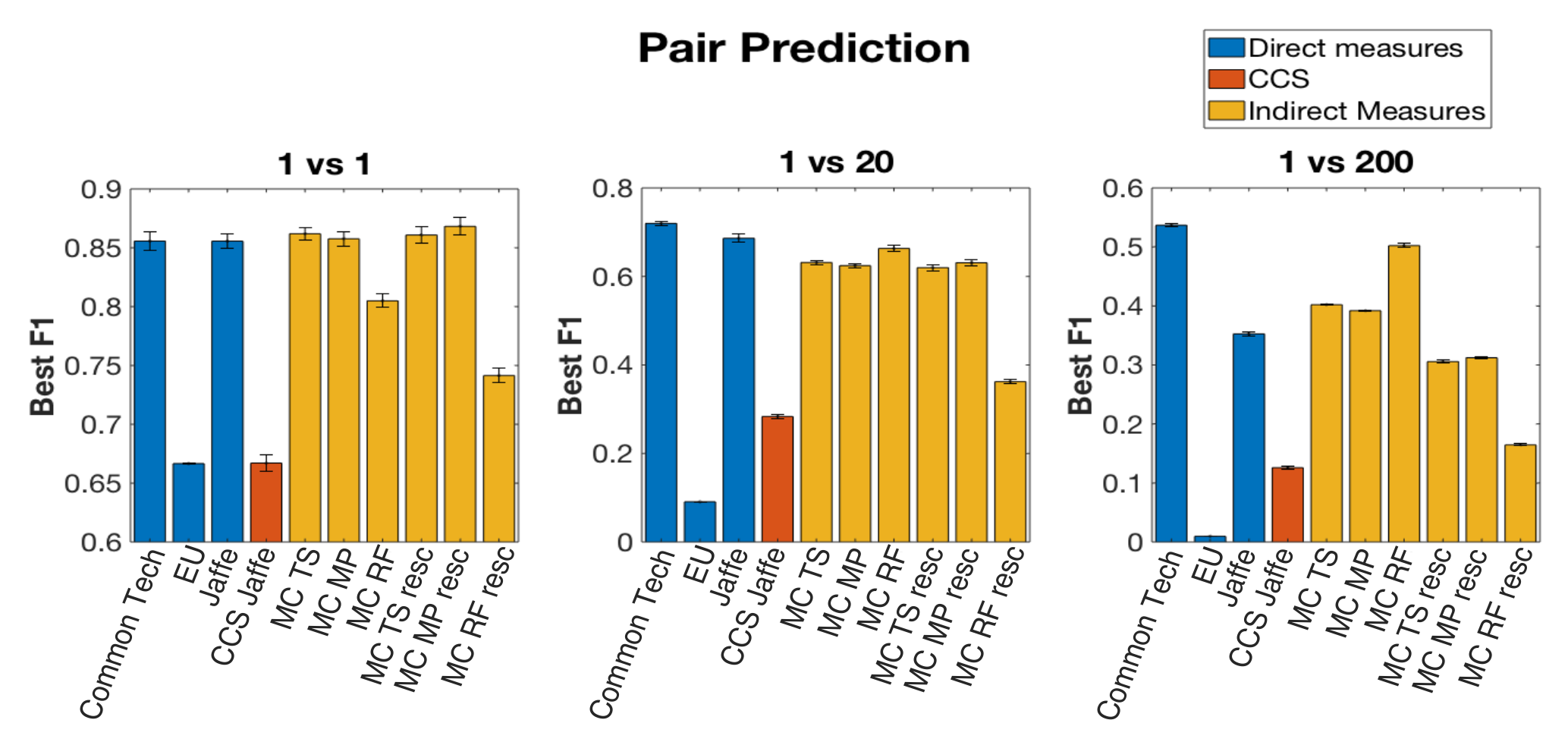}
\caption{\textbf{Comparison between Best F1 on Pair Prediction.} We show the comparison between our defined similarity metrics. We chose as values of Class Imbalance: 1 VS 1, 1 VS 20, and 1 VS 200, which means that for each true M\&A we extract 1, 20, and 200 false negatives, i.e. 1, 20, or 200 not real M\&A. In both figures, we use three different colors for each different category's metrics. Blue represents the direct metrics, yellow the indirect ones and orange the CCS ones. Error bars are present because for each class imbalance we repeat the exercise of selecting random Target companies and/or Acquirer companies $20$ times, calculating mean and standard deviation of the mean. In this case, it was not possible to construct the Jaffe + Sectors measure and the relative CCS because we don't have information on the industrial Sector of all companies.}
\label{fig:bf1_PP_all}
\end{center}
\end{figure}
This is the case of indirect measures based on networks and Random Forest and the direct measures based on networks. The diversification of companies in our database follows a power-like distribution, where low diversified firms are prevalent (see Figure \ref{fig:DistPat}). This implies that, if the negative examples companies are drawn from the entire database, using metrics which is correlated with diversification leads to better predictions.
In Figure \ref{fig:bf1_PP_all} we present the Best F1 values relative to the Pair Prediction task in this case.
Note that with these conditions, it was not possible to construct the Jaffe + Sectors measure and the relative CCS because we don't have information on the industrial Sector of all companies.
As it is possible to see in Figure \ref{fig:bf1_PP_all}, all the rescaled metrics perform worse than the original ones. This is true also for the Jaffe - Common Tech pair of measures in facts, now the best performing measure for predictions is Common Tech. 
Note that in any case, we found that a direct measure based on an angular distance between companies in the technology space is better than any other direct or indirect metric employed.
As a final remark we can note that independently from the relative Best F1 levels among the various measures, predictions done using negative examples from the whole database are much better than the ones presented in the Result section, done extracting the negative examples from a set of companies whose industrial sectors are known.
This demonstrates, in general, that the choice of negative examples strongly influences the power of our predictions.

\section{Evaluation metrics robustness test}
\begin{figure}[h!]
\centering
\subfloat%[]%[\emph{.}]
   {\includegraphics[width=.47\textwidth]{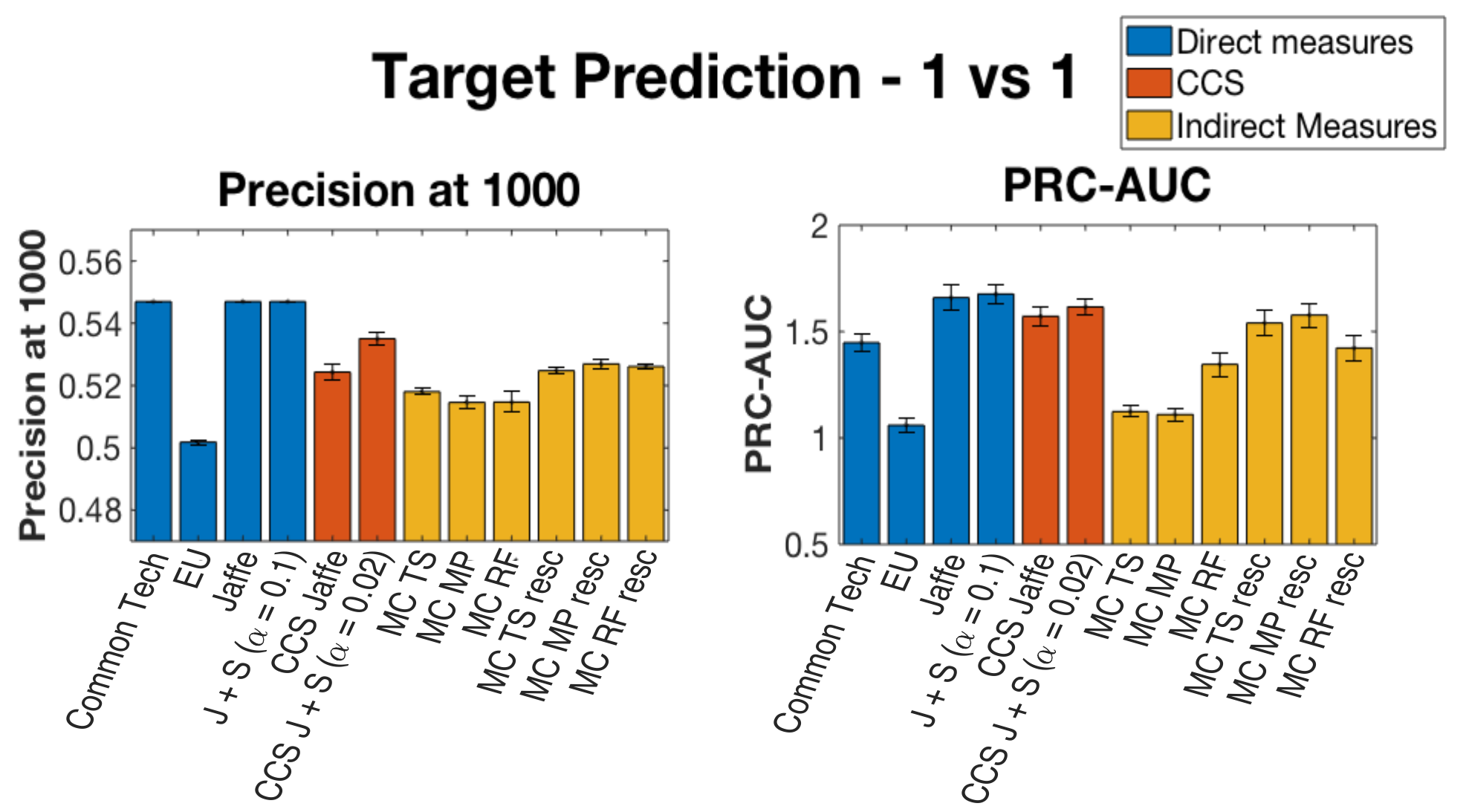}} 
\subfloat%[]%[\emph{.}]
   {\includegraphics[width=.47\textwidth]{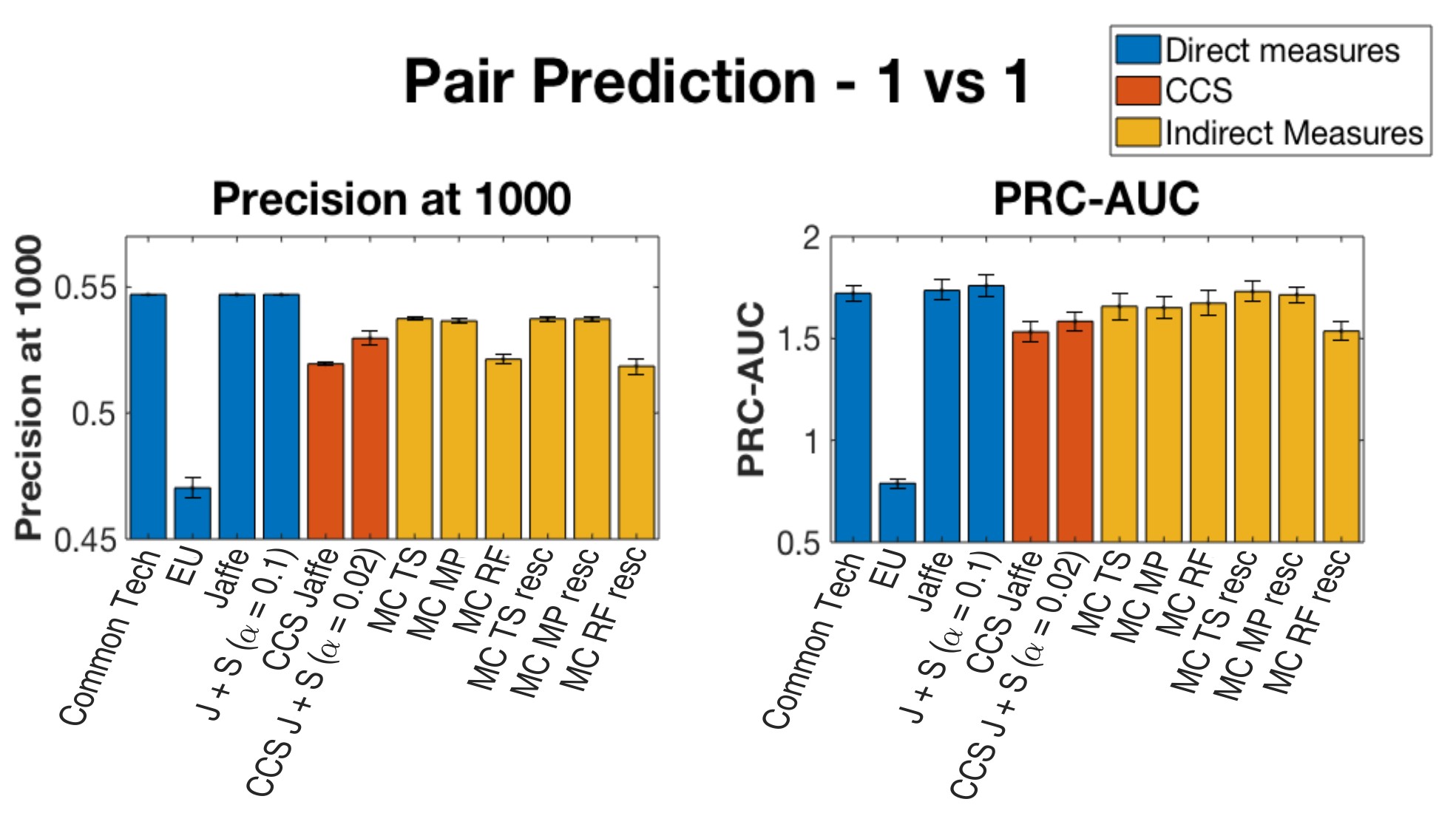}} \\
\subfloat%[]%[\emph{.}]
   {\includegraphics[width=.47\textwidth]{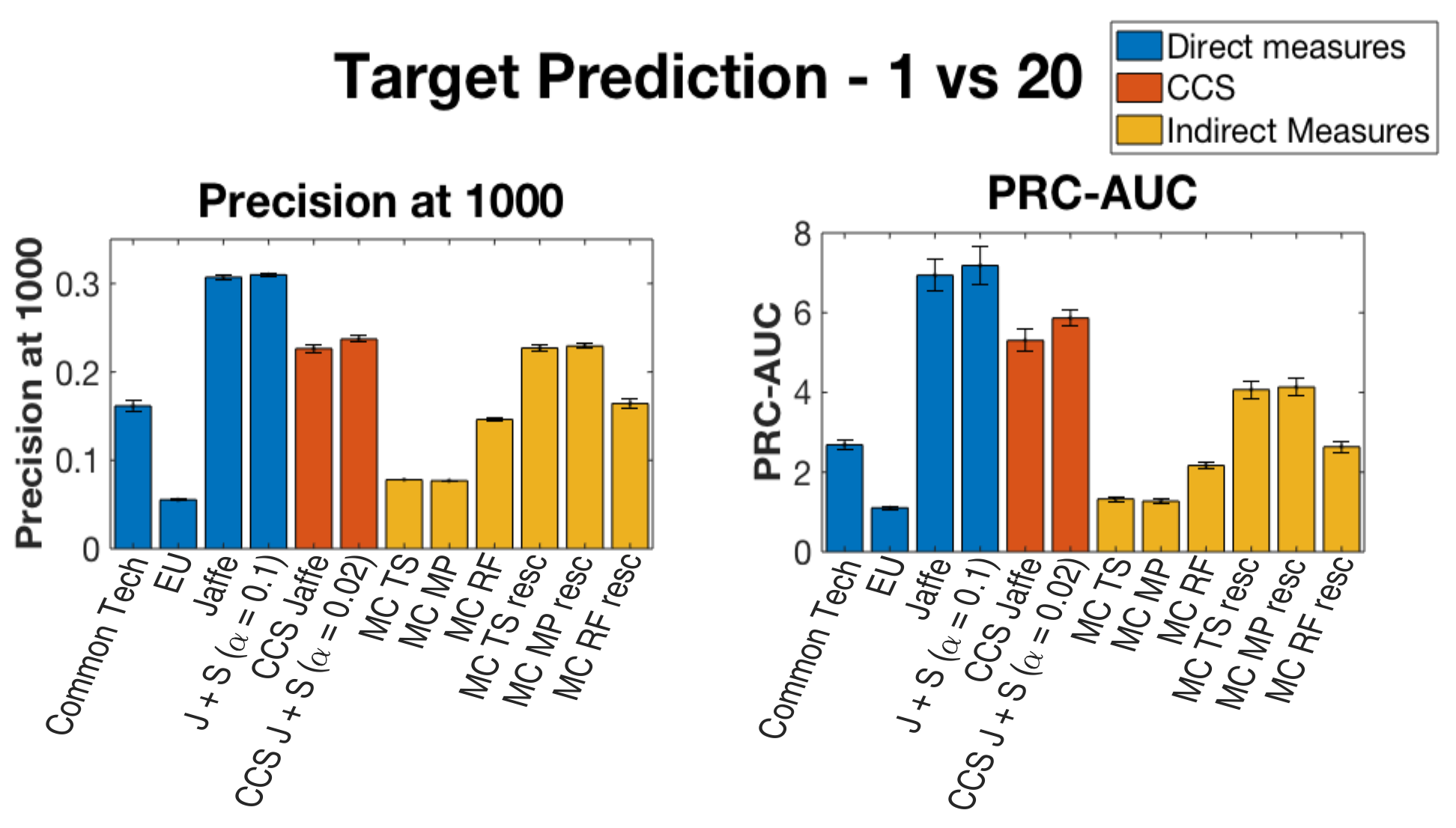}} 
\subfloat%[]%[\emph{.}]
   {\includegraphics[width=.47\textwidth]{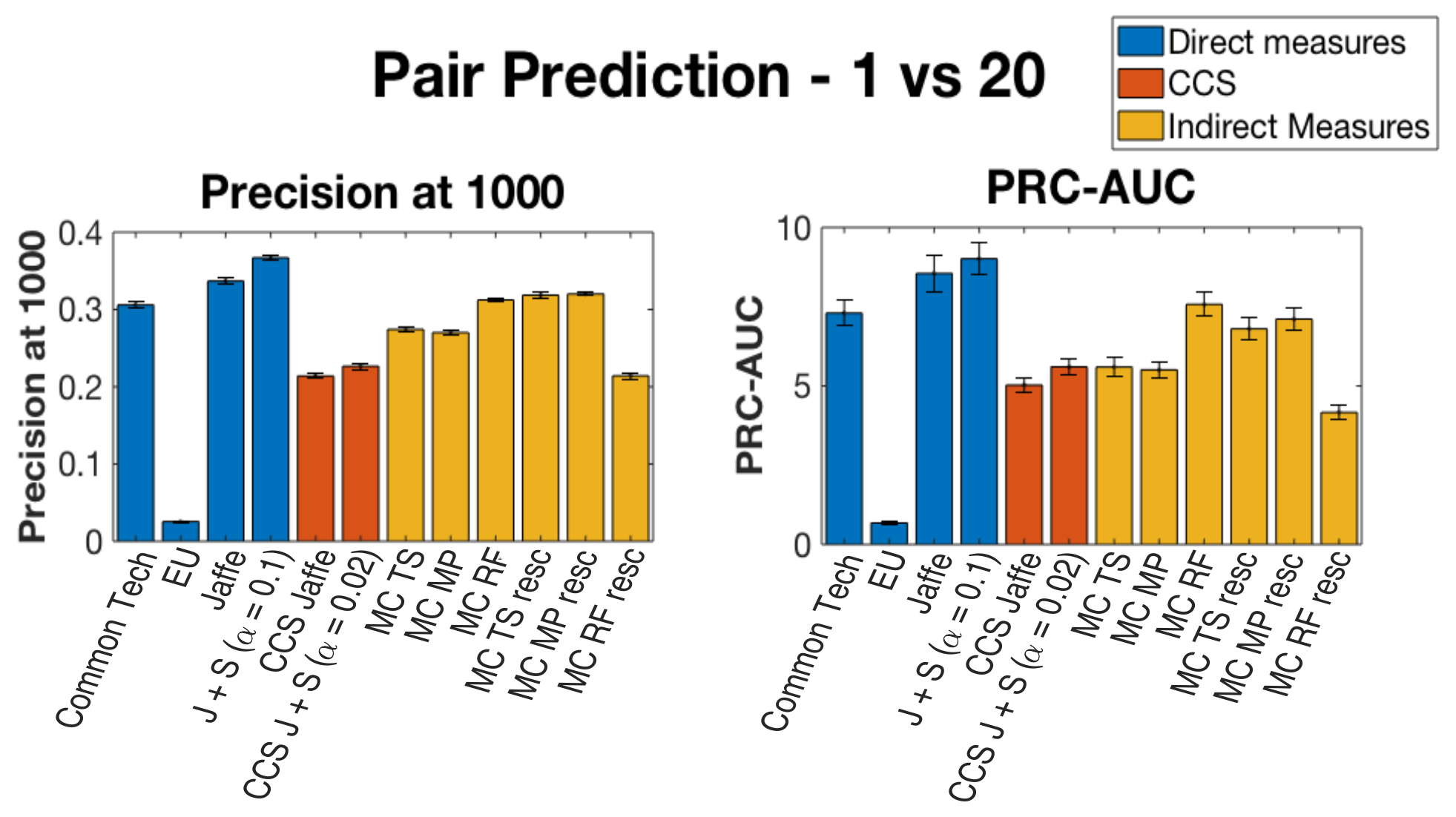}} \\
\subfloat%[]%[\emph{.}]
   {\includegraphics[width=.47\textwidth]{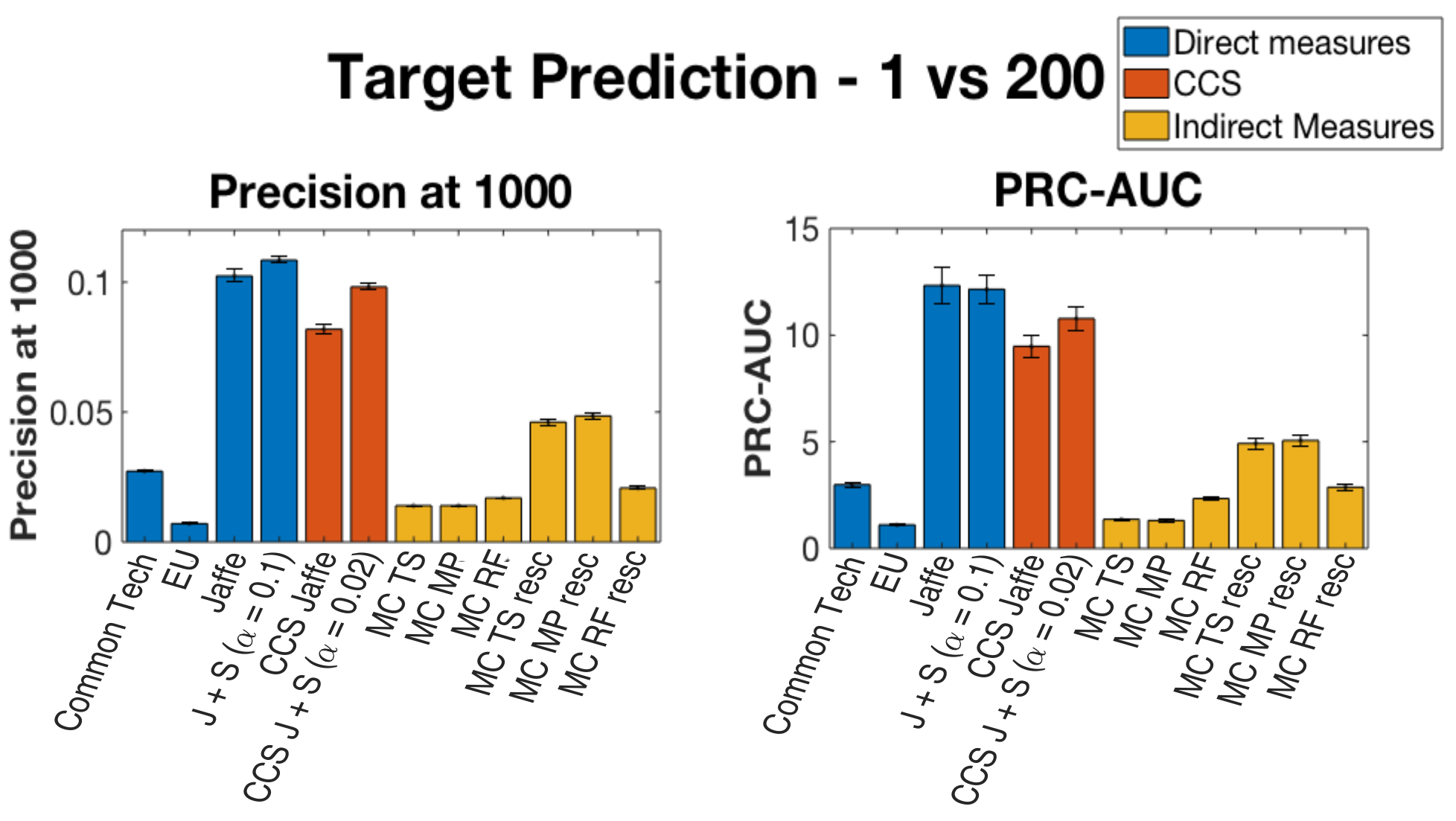}}
\subfloat%[]%[\emph{.}]
   {\includegraphics[width=.47\textwidth]{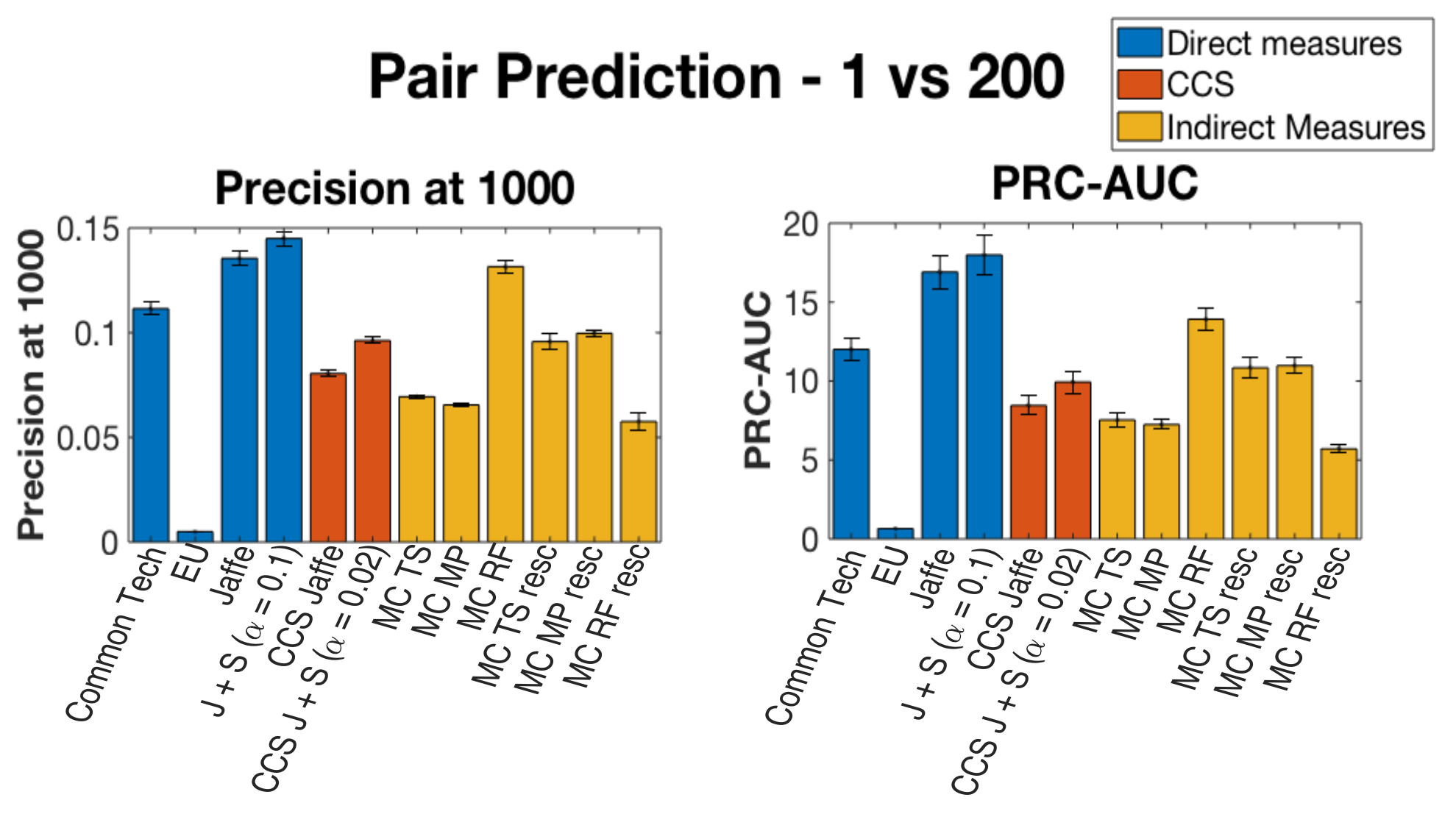}} 
\caption{\textbf{Predictions evaluation of Target and Pair forecasts for other metrics.} In both figures, we show the comparison between our defined similarity metrics. We chose as values of Class Imbalance: 1 VS 1, 1 VS 20, and 1 VS 200, which means that for each true M\&A we extract 1, 20, and 200 false negatives, i.e. 1, 20, or 200 not real M\&A. In both figures, we use three different colors for each different category's metrics. Blue represents the direct metrics, yellow the indirect ones and orange the CCS ones. Error bars are present because for each class imbalance we repeat the exercise of selecting random Target companies and/or Acquirer companies $20$ times, calculating mean and standard deviation of the mean. }
\label{fig:aoao}
\end{figure}
In this Section we reproduce the results of Figure 2 in the main text, using different evaluation metrics. The metrics we decided to use are:
\begin{itemize}
    \item Area Under Precision-Recall Curve (PRC-AUC) \cite{beger2016precision}: the area under the curve in the precision-recall plane. The latter quantities are obtained by varying the threshold that identifies the value above which the scores are associated with positive predictions. We decide to calculate it normalizing by the random case in the same way of Futagami et al. work \cite{FUTAGAMI202122}.
    %\item Area Under ROC Curve (ROC-AUC) \cite{narkhede2018understanding}: the area under the curve in the True Positive Rate (TPR) and False Positive Rate (FPR) plane. It represents the probability that a randomly selected positive instance will receive an higher score than a randomly selected negative instance.
    \item Precision at 1000: the fraction of the largest $1000$ elements of the score predictions that are actually true.
\end{itemize}
%For completeness, we also report the results obtained with the Best-F1 of the main.
The results, Figure \ref{fig:aoao}, are consistent with those in the main text using the Best-F1 metric with the Jaffe + Sector that is the best prediction algorithm. About the best $\alpha$ for both Jaffe + Sector and CCS J + S, the best one is the same of the Best-F1 metric.

\section{Coherence $\gamma$ and $RF$ VS diversification}
\begin{figure}[h!]
\centering
\subfloat[]%[\emph{.}]
   {\includegraphics[width=.4\textwidth]{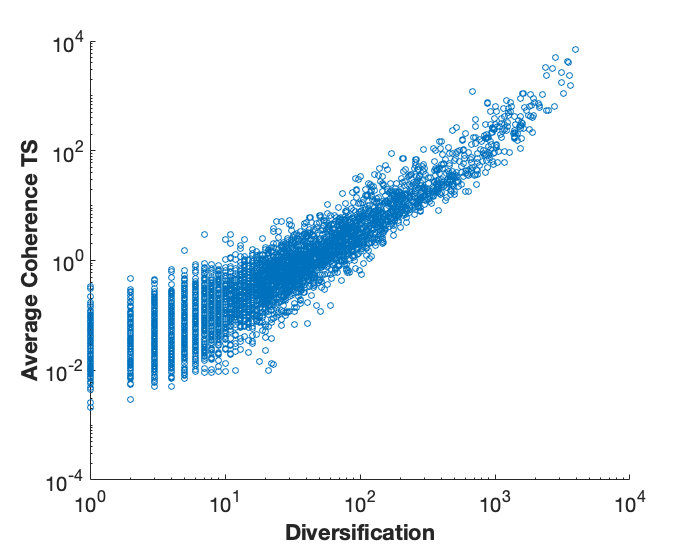}} 
\subfloat[]%[\emph{.}]
   {\includegraphics[width=.4\textwidth]{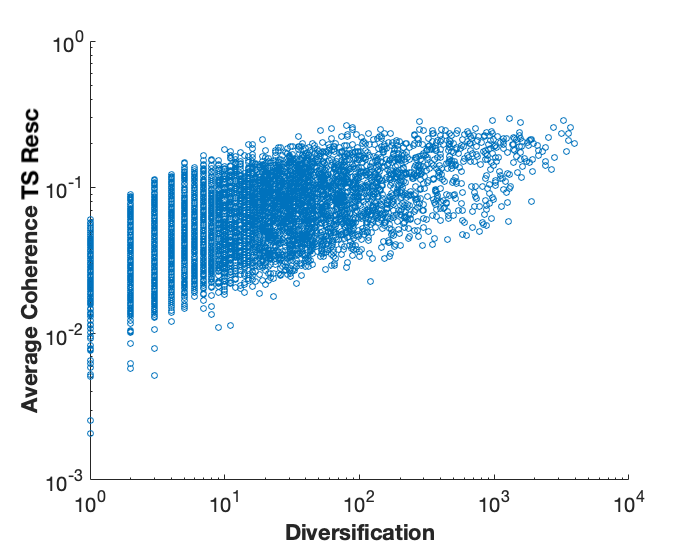}} \\
\subfloat[]%[\emph{.}]
   {\includegraphics[width=.4\textwidth]{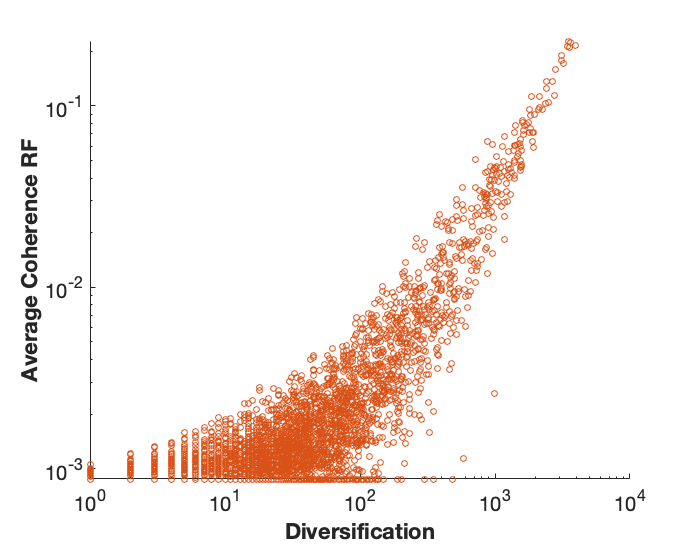}} 
\subfloat[]%[\emph{.}]
   {\includegraphics[width=.4\textwidth]{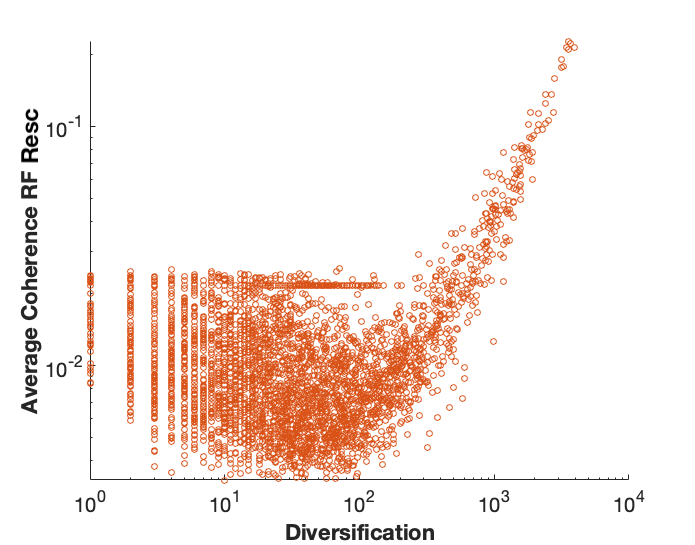}}
\caption{\textbf{Indirect measures VS diversification: before and after rescalation} In \textbf{a} and \textbf{c} we show the correlation between the Coherence $\gamma$ and $RF$, and the diversification before the rescalation process. To quantify the correlation between these quantities, we compute the spearmanr correlation, equal to $0.90$ for $\gamma$ and $0.69$ for $RF$. After the rescalation process, in \textbf{b} and \textbf{d} we show the new correlation, with a spearmanr correlation of $0.65$ for $\gamma$ and $0.42$ for $RF$. The rescalation can lower the correlation between the two quantities, and the effects of this can be seen in the main text Results.}
\label{fig:div}
\end{figure}
In the main text, we explain the necessity to rescale the Indirect measure. The reason is because $\gamma$ and $RF$ are highly correlated with the diversification. We can see these correlations in Figure \ref{fig:div}\textbf{a}-\textbf{c}. Rescaling between $0$ and $1$ these measures, we can partially remove that dependence (as seen in Figure \ref{fig:div}\textbf{b}-\textbf{d}) and quantify its effects on the forecasts.\\
To quantify the correlation between these quantities the we compute the absolute value of the spearmanr correlation, equal to $0.90$ for $\gamma$ and $0.69$ for $RF$. After the rescalation process, in \textbf{b} and \textbf{d} we show the new correlation, with a spearmanr correlation of $0.65$ for $\gamma$ and $0.42$ for $RF$. The rescalation can lower the correlation between the two quantities, and the effects of this can be seen in the main text Results.

\section{Optimisation of $\alpha$ in CCS Jaffe + Sector}
\begin{figure}[h!]
\begin{center}
\includegraphics[scale=.80]{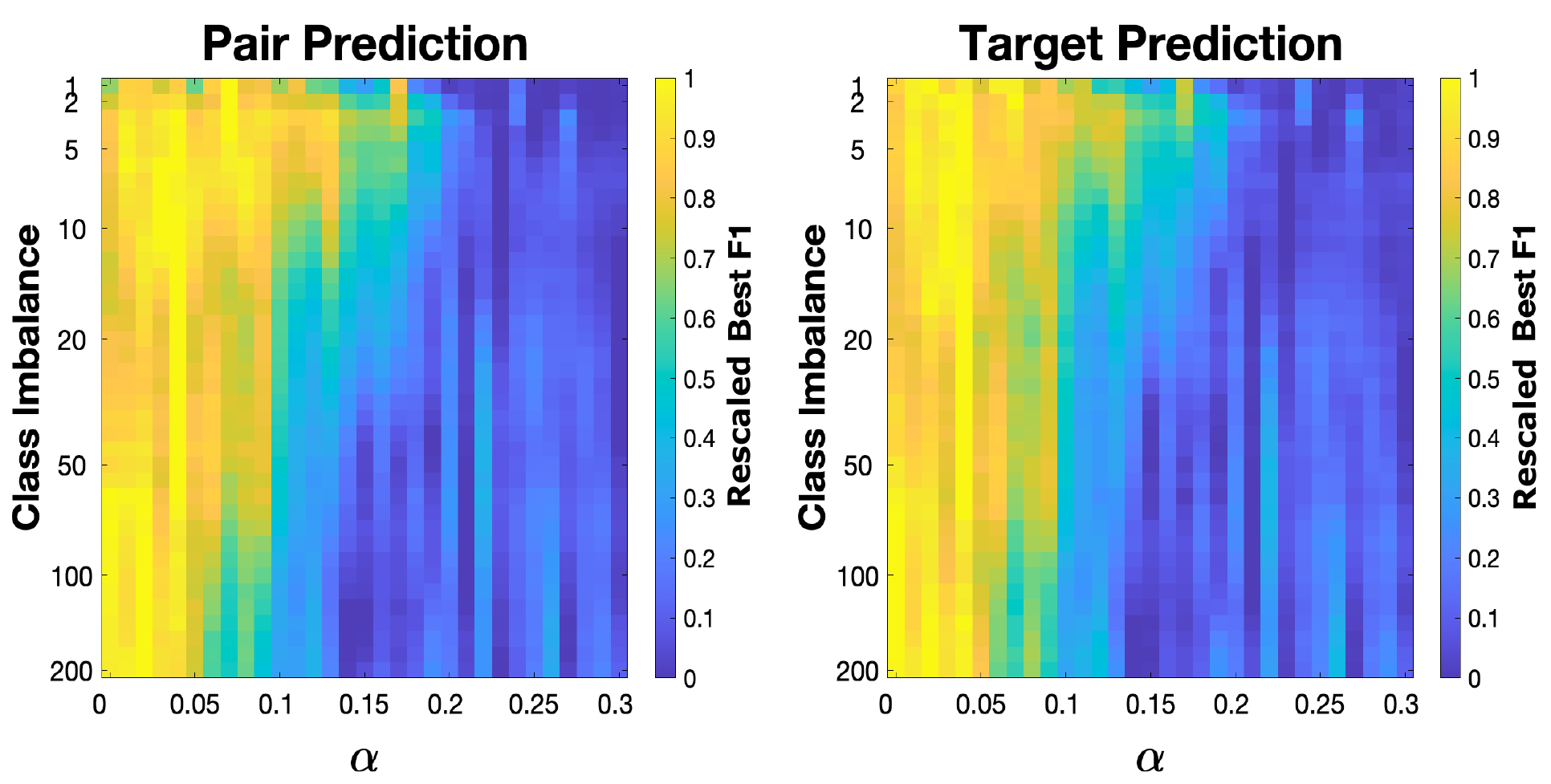}
\caption{\textbf{Dependence of maximum Best F1 on $\alpha$ and Class imbalance in CCS Jaffe + Sectors model}. We show the behavior of the Best F1 for $\alpha$ and the class imbalance. For each value of class imbalance, we rescale the Best F1 between 0 and 1 to better see the maximum as a function of $\alpha$. The Best F1's maximum remains around low $\alpha$ values for different values of class imbalance, and this is becomes more evident to grow of the class imbalance. This suggests that when choosing a M\&A pair among a large pool of options, the similarity between the companies from the point of view of the industrial sector becomes increasingly less important in contrast to a technological similarity.}
\label{fig:opt_CCS}
\end{center}
\end{figure}
To find the best $\alpha$ parameter for the CCS measure, we do the same computations explained in the Methods Section main text in the Jaffe case. Unlike the optimization of $\alpha$ over Jaffe, in the case of CCS we notice that starting from relatively low values of class imbalance (1VS5, for example), the addition of industry information leads to lower forecast results. In other words, starting from such values of class imbalance the optimal $\alpha$ is close to $0$. In Figure \ref{fig:opt_CCS} we show the heat map of the Best F1 as a function of $\alpha$ and the class imbalance, both for the Pair Prediction and the Target Prediction. To average out some fluctuations coming from the random extraction of negative examples, each point in the Figures reports the mean Best F1 over 20 realizations of the predictions.
Because the Best F1 is highly correlated with the class imbalance, in these Figures, for each value of $N$ we rescaled the Best F1 between 0 and 1. In this way, it is possible to better spot the maximum of Best F1 as a function of $\alpha$, for each value of class imbalance.
As it is possible to extrapolate from the figures, if we increase the class imbalance, the maximum of Best F1 remains around low $\alpha$ values for different values of class imbalance, becoming increasingly evident as class imbalance increases. This suggests that when choosing a M\&A pair among a large pool of options, the similarity between the companies from the point of view of the industrial sector becomes increasingly less important in contrast to a technological similarity.

\section{Crunchbase name matching}
To associate Cruchbase companies with BVDIDs and then with their technological portfolios, it was necessary to match their names to those contained within AMADEUS, in which the BVDID of the companies are reported.
To get a good match between the names, these were "cleaned" and standardized, performing a removal of symbols, special characters, spaces and company acronyms (S.p.a, srl, etc.). For this last task, the \textit{cleanco} package for Python was used.
In general, in the names cleaning process we followed these steps:
\begin{enumerate}
	\item Conversion of names to uppercase, removal of commas, dashes and text between parentheses.
	\item First execution of \textit{cleanco}, dots removal and second execution of \textit{cleanco}, removal of European nations' names.
	\item Removal of all remaining symbols, special characters and spaces.
\end{enumerate}
After the cleaning, the names match was performed for string equality. 
To limit the errors due to poor or excessive cleaning of names, the match was performed 3 times with 3 different levels of cleaning and the results were then merged.
The 3 cleaning levels differ in the partial or total execution of the second step: in the first level \textit{cleanco} is executed only once, in the second level the  second execution of \textit{cleanco} takes place after the removal of dots and, finally, in the third level names of the European nations are removed.
In this way, it was possible to associate at least a BVDID to 28137 companies out of the 123576 in the Crunchbase database and involved in acquisitions.

\section{Sectors Classification}
The companies in the Crunchbase database are organized, according to the industrial sector, on two levels of aggregation: the lowest counts 744 \textit{categories}, while the highest counts 43 \textit{category groups}. The classification is not related to the official ones (NACE NAICS, SIC etc.), but was built independently by Crunchbase.
As it is constructed, this classification is not useful for the purposes of this work because:
\begin{itemize}
\item Many \textit{category groups} (and consequently several \textit{categories}) are associated with a single company.
\item The \textit{category groups} number is too large for visualisation purposes.
\end{itemize}
For these reasons, starting from the classification of Crunchbase in \textit{category groups}, we defined a new classification in 13 industrial sectors, identified by the letters from A to M.
In this way we are able to assign a unique industrial sector to about 70\% of the companies that we had previously linked to their technological portfolio.

The construction of this classification took place in 4 phases:
\begin{itemize}
\item First division of \textit{category groups} into 12 sectors.
\item Assignment of sectors A and B \textit{Manufacturing and Goods} and \textit{Hardware  \& Software} to companies: the associated \textit{category groups} are the most common, therefore only the companies which were assigned exclusively to the respective \textit{category group} have been classified in sectors A and B.
\item Assignment of remaining sectors: depending on the associated \textit{category groups} one or more sectors have been assigned to each company.
\item Specific reassignments: some companies, linked to two sectors, have been reassigned to only one of the two. In this phase, the thirteenth sector, the \textit{Medical Research} one, was created.
\end{itemize}
Sectors, associated \textit{category groups} and subsequent reassignments are listed in Table \ref{tab:Sectors}.

%\begin{table}[htbp]
%\centering
%\caption{Industrial sector classification for Crunchbase's companies}
\newpage
\fontsize{10}{10}\selectfont
\begin{longtable}{llll}
\caption{Industrial sector classification for Crunchbase's companies}\\
\label{tab:Sectors}

 %\noalign{\smallskip} \hline \hline \noalign{\smallskip}
\textbf{Sectors} & \textbf{Label} & \textbf{Reassigned Labels} & \textbf{Category Groups} \\
   \hline \noalign{\smallskip}
 
Manufacturing \& Goods & A     &       & Consumer Goods \\
      &       &       & Manufacturing \\
\hline
Hardware \& Software & B     &       & Hardware \\
      &       &       & Software \\
\hline
Clothing \& Food & C    &   CG   & Agriculture and Farming \\
      &       &       & Clothing and Apparel\\
      &       &       & Food and Beverage\\
\hline
Software \& Gaming &D    & DF, DL, DE & Apps\\
      &       &       & Consumer Electronics\\
      &       &       & Gaming\\
      &       &       & Mobile\\
      &       &       & Navigation and Mapping\\
      &       &       & Platforms\\
\hline
Research & E     &       & Artificial Intelligence\\
      &       &       & Biotechnology\\
      &       &       & Data and Analytics\\
      &       &       & Science and Engineering\\
\hline
Energy \& Transport & F     & EF, DF    & Energy\\
      &       &       & Natural Resources\\
      &       &       & Sustainability\\
      &       &       & Transportation \\
\hline
Community Services & G    &      & Community and Lifestyle\\
      &       &       & Design\\
      &       &       & Education\\
      &       &       & Events\\
      &       &       & Media and Entertainment\\
      &       &       & Real Estate\\
      &       &       & Sports\\
      &       &       & Travel and Tourism\\
\hline
Services \& Security & H     & HL    & Administrative Services\\
      &       &       & Government and Military\\
      &       &       & Privacy and Security\\
      &       &       & Professional Services\\
\hline
Health Care &  I     &       & Health Care\\
\hline
Economy \& Finance & J     &       & Financial Services\\
      &       &       & Lending and Investments\\
      &       &       & Payments\\
\hline
Marketing & K     & KL, GK & Advertising\\
      &       &       & Commerce and Shopping\\
      &       &       & Content and Publishing\\
      &       &       & Sales and Marketing\\
\hline
ICT   & L     &    EL   & Information Technology\\
      &       &       & Internet Services\\
      &       &       & Messaging and Telecommunications\\
      &       &       & Music and Audio\\
      &       &       & Video\\
\hline
Medical Research & M     & EI    &  \\

\hline
\end{longtable}

%\end{table}

\bibliography{sample}